\documentstyle[12pt,epsfig]{article}
\newcommand{\be}{\begin{equation}}
\newcommand{\ee}{\end{equation}}
\newcommand{\bea}{\begin{eqnarray}}
\newcommand{\eea}{\end{eqnarray}}

\catcode`@=11
\newcount\@tempcntc
\def\@citex[#1]#2{\if@filesw\immediate\write\@auxout{\string\citation{#2}}\fi
  \@tempcnta\z@\@tempcntb\m@ne\def\@citea{}\@cite{\@for\@citeb:=#2\do
    {\@ifundefined
       {b@\@citeb}{\@citeo\@tempcntb\m@ne\@citea\def\@citea{,}{\bf ?}\@warning
       {Citation `\@citeb' on page \thepage \space undefined}}%
    {\setbox\z@\hbox{\global\@tempcntc0\csname b@\@citeb\endcsname\relax}%
     \ifnum\@tempcntc=\z@ \@citeo\@tempcntb\m@ne
       \@citea\def\@citea{,}\hbox{\csname b@\@citeb\endcsname}%
     \else
      \advance\@tempcntb\@ne
      \ifnum\@tempcntb=\@tempcntc
      \else\advance\@tempcntb\m@ne\@citeo
      \@tempcnta\@tempcntc\@tempcntb\@tempcntc\fi\fi}}\@citeo}{#1}}
\def\@citeo{\ifnum\@tempcnta>\@tempcntb\else\@citea\def\@citea{,}%
  \ifnum\@tempcnta=\@tempcntb\the\@tempcnta\else
   {\advance\@tempcnta\@ne\ifnum\@tempcnta=\@tempcntb \else \def\@citea{--}\fi
    \advance\@tempcnta\m@ne\the\@tempcnta\@citea\the\@tempcntb}\fi\fi}
\catcode`@=12

\begin{document}

\titlepage

\begin{flushright}
MC-TH-98/6 \\
CERN-TH-98-275 \\
Liverpool LTH 431
\end{flushright}
\centerline{{\Large {\bf The Massive Yang-Mills Model}}}
\centerline{{\Large {\bf and Diffractive Scattering}}}
\vskip 0.8cm
\centerline{\bf{ J.R.\ Forshaw$^{a}$, J.\ Papavassiliou$^{b}$,
C.\ Parrinello$^{c}$}}
\vskip 0.2cm
\centerline{$^{a}$ Department of Physics \& Astronomy}
\centerline{University of Manchester}
\centerline{Manchester M13 9PL, U.K.}
\vskip 0.2cm
\centerline{$^b$ Theory Division, CERN} 
\centerline{CH-1211 Geneva 23}
\centerline {Switzerland.}
\vskip 0.2cm
\centerline{$^c$ Department of Mathematical Sciences}
\centerline{University of Liverpool}
\centerline{Liverpool L69 3BX, U.K.}
\begin{abstract}
  We  argue that  the massive  Yang-Mills  model of  Kunimasa \& Goto,
  Slavnov,  and  Cornwall, in which   massive gauge vector  bosons are
  introduced  in a gauge-invariant way  without resorting to the Higgs
  mechanism, may  be useful for studying  
  diffractive  scattering 
  of strongly interacting particles.
  With this motivation, we perform in this model explicit calculations
  of  $S$-matrix elements  between quark states,  at   tree level, one
  loop, and  two loops, and   discuss issues of  renormalisability and
  unitarity.  In particular, it is  shown that the $S$-matrix  element
  for  quark  scattering is renormalisable at  one-loop order, and is
  only  logarithmically    non-renormalisable  at two     loops.  
  The  discrepancies in the  ultraviolet regime    between  the  one-loop
  predictions of this model and those of massless  QCD are discussed in
  detail.    In addition, some  of  the  similarities and  differences
  between the massive  Yang-Mills  model  and  theories with  a  Higgs
  mechanism  are  analysed  at the  level  of the  $S$-matrix.   
  Finally, we briefly discuss the high-energy behaviour of the leading 
  order amplitude for quark-quark elastic scattering in the 
  diffractive region.  
The above
  analysis sets up the stage for carrying out a systematic computation
  of  the higher order corrections  to the two-gluon exchange model of
  the Pomeron using massive gluons inside quantum loops.

\end{abstract}

\newpage

\section{Introduction}

The quantitative   description of  diffractive  phenomena  within  the
framework of QCD is a long-standing problem.   In part, the difficulty
arises because diffractive processes involve both hard and soft scales
resulting  in a   complicated    interplay between perturbative    and
non-perturbative   effects.  One  way  to tackle   this problem is  to
attempt a description using a ``dressed''  version of the perturbative
degrees of  freedom, where the ``dressing" is  meant to mimic the role
of  non-perturbative effects.   Following Low  \cite{Low} and Nussinov
\cite{Nussinov}, Landshoff  and Nachtmann  (LN) \cite{LN} introduced a
two-gluon exchange-model of diffractive scattering, where they assumed
that the  infrared  behaviour of the   (abelian)  gluon propagator  is
modified by   non-perturbative effects.  Their  success in reproducing
several of the  features  of Pomeron  exchange suggests  that such  an
attempt may  not be totally  futile, and makes the  question of how to
compute systematically higher order corrections  within this model all
the more interesting.  \footnote{ A different  approach is provided by
  the BFKL formalism \cite{BFKL}, which is the most serious attempt at
  a first  principles  QCD derivation of  Pomeron  exchange to   date. 
  However, the perturbative nature of the BFKL approach often makes it
  unsuitable for the  analysis of  diffractive scattering, where  both
  soft and hard momentum scales are in general relevant.}

In  the LN picture  of the  Pomeron the need  for  modifying the gluon
propagator  arises as follows: The simplest  Feynman diagram which can
model the Pomeron (exchange carrying the quantum
numbers of the vacuum) is a box-diagram where two off-shell gluons are
exchanged between incoming  and outgoing quarks,  which  scatter
elastically.  The perturbative calculation of  the above process gives
rise to  an  infrared singularity at  $t=0$, whose  origin is the fact
that the  bare  gluon  propagator $d_0  (q^2)$  diverges at  $q^2=0$.  
Specifically,  the amplitude obtained  from such a diagram assumes the
form  $i\beta_0^2 (\bar{u}\gamma_{\mu}u)(\bar{u}\gamma^{\mu}u)$, where
$\beta_0^2$ is given by  $\beta_0^2 \sim \int dq^2 [\alpha_s  d
(q^2)]^2$, and $d  (q^2)$ is the  gluon propagator. The introduction of
a ``massive'' gluon  propagator is  the simplest  way to
obtain a finite $\beta_0^2$ and the gluon mass is then fixed by data.

It   has  been  suggested long ago  \cite{glmassJMC1}    that the
non-perturbative dynamics of QCD lead to the generation of a dynamical
gluon mass \footnote{Dynamically generated masses depend non-trivially
  on the momentum; in particular, they vanish for large momenta.  This
  property   is   crucial  for  the  renormalisability  of  the theory
  \cite{runmass}.}  while the local   gauge invariance of the   theory
remains   intact.  This gluon  ``mass''  is  not a directly measurable
quantity, but must be related to other physical parameters such as the
string tension, glueball masses, or  the QCD vacuum energy \cite{SVZ},
and furnishes, at   least in principle,  a regulator  for all infrared
divergences of QCD.  The  above picture emerged  from  the study  of a
gauge-invariant set  of Schwinger-Dyson equations  \cite{glmassJMC2}.  
In  addition, lattice computations  \cite{lattice} reveal the onset of
non-perturbative effects which can be modelled by means of effectively
massive gluon  propagators.   Various  independent  field  theoretical
studies   spanning  almost two  decades \cite{ZwanzigerGribov,various}
also corroborate some type of  mass generation, although no  consensus
about  the  exact nature  of the   mass-generating  mechanism has been
reached thus far \footnote{Of course, the introduction of a gluon mass
  at tree  level through the   usual Higgs  mechanism \cite{Higgs}  is
  excluded, as it would introduce extra (unwanted) scalar particles in
  the physical  spectrum.}.  Interestingly enough, the effective gluon
propagator   derived    in  \cite{glmassJMC2} describes    successfully
nucleon-nucleon scattering when  inserted, in a  rather heuristic way,
into the two-gluon exchange  model \cite{diffraction2}.   Despite this
phenomenological success however, it  is not clear whether a dynamical
gluon  propagator may   be  used  in calculations  as   if it  were  a
tree-level propagator derived from Feynman  rules. More
importantly, it is not known how to systematically improve upon such a
calculation, i.e. how to compute higher order corrections.

Given this lack  of a computational  scheme originating from a ``first
principles''  QCD treatment, we  propose instead to  resort to a field
theory which is  formally close to  QCD and contains  at the same time
the feature which  appears  to be phenomenologically useful,   namely a
gluon  mass.  To that end we  revisit a model introduced independently
by Kunimasa  \&   Goto \cite{Kunimasa},  Slavnov   \cite{Slavnov}, and
Cornwall   \cite{Cornwall},  which    provides the  extension   to   a
non-Abelian     context   of  the  work   of  Stueckelberg
\cite{Stueckelberg}.  This  model accommodates  massive vector  bosons
{\it without} compromising local gauge   invariance and {\it  without}
introducing a Higgs  sector.  In what  follows we will  refer to it as
the massive Yang-Mills (MYM) model.

In the MYM model, a mass term is added directly to the Yang-Mills (YM)
Lagrangian and gauge   invariance   is preserved  with   the  help  of
auxiliary  scalar    fields.  Unlike     the usual    Higgs  mechanism
\cite{Higgs}  however, there are  no additional physical   particles
appearing in  the spectrum (no Higgs bosons).   The price  one pays is
that perturbative    renormalisability   is    lost.
In particular, the one-loop $S$-matrix element for gluon
elastic  scattering $gg\to  gg$   is known   to  be non-renormalisable
\cite{nonren1}; its renormalisability can  be restored only  with the
introduction of Higgs boson in the  spectrum \cite{CLT,CLS,LQT}.  This
fact renders the  MYM  non-renormalisable at one loop.   However,  the
introduction  of  a   Higgs boson  is  {\sl   not}  necessary  for the
renormalisability   of the one-loop  $S$-matrix  of  the processes $q
\bar{q} \to q \bar{q}$ which is relevant for diffractive
scattering. As we will see in detail, the first time this
latter process receives (logarithmically)
non-renormalisable contributions is at two loops. 
In addition, the model  has been  shown to be
unitary, in the sense  
of the optical theorem, to all orders in
perturbation theory \cite{Slavnov}.
Several formal properties of  this model have been extensively studied
in the   literature cited above  and are  well understood.

Our main phenomenological 
motivation for turning  to the MYM model  is to carry out the
next-to-leading order corrections to the two-gluon exchange process 
for $q
\bar{q} \to q \bar{q}$, in
the context of a concrete  field theory, where the effects originating
from  the presence of  a gluon  mass can  be  studied systematically.  
Clearly, before attempting such a complex calculation it is necessary to 
develop some familiarity  
with the  predictions of the MYM model at leading and next-to-leading 
order. 
  The purpose of the  present work is  to provide a detailed
analysis of  various  field-theoretical issues which appear   when one
uses the  MYM model for   computing $S$-matrix elements \footnote{  By
  working directly with  $S$-matrix  elements one has  the  additional
  advantage    of   avoiding  pathologies    which affect  individual,
  unphysical    Green's  functions.    In   fact,  because of  several
  cancellations taking place at the  level of $S$-matrix elements, the
  final answer  often has better properties than  those of the Green's
  functions  involved in the calculation.  A  typical  example of this
  situation arises when  using the unitary  gauges for the electroweak
  model; in  these gauges,  Green's functions  are non-renormalisable,
  while $S$-matrix  elements are  \cite{HMren}.}  involving quarks  as
external  states.   In addition to  the  clarification  of theoretical
points,  several of the  results  presented  in this paper  constitute
useful ingredients of the full calculation.
  
More specifically, we discuss the following points: 
\begin{itemize}
\item We  analyse  in detail how the   MYM and QCD   differ already at
  tree-level, and how this difference propagates to higher orders.  In
  particular we show using both unitarity and analyticity arguments as
  well    as explicit  one-loop     calculations  how the   tree-level
  discrepancy affects  the one-loop beta function,  i.e. it alters the
  high energy behaviour of the theory.
  
\item We  verify explicitly in the  context of a specific example that
  the S-matrix contains no unphysical poles.  The cancellation of such
  poles,  which  is expected  from formal considerations,  provides  a
  non-trivial consistency  check of  the  model, and  can  serve as  a
  guiding principle when carrying out lengthy calculations.
  
\item  We demonstrate  that at   the   one-loop level  the  scattering
  amplitude of interest is renormalisable,  and that one can construct
  a  gauge-invariant running  coupling (effective  charge)  just as in
  QCD.   This leads   to  the definition   of a  gauge-invariant gluon
  propagator,   generalising Cornwall's construction  for the standard
  QCD case.  A detailed comparison  of our result  with the QCD one is
  performed.
 
\item We show that the non-renormalisable  contribution arising at the
  two-loop  level  depends  only logarithmically  on the  cutoff. This
  result is new, to the  best of our  knowledge; its derivation relies
  crucially  on extensive cancellations which take  place at the level
  of the $S$-matrix after the judicious exploitation of the tree-level
  Ward identities of the MYM.

\end{itemize}
The paper is organised  as follows: In Section  \protect\ref{sec:mqcd}
we briefly  review the MYM  formalism, and establish connections which
will be  useful for the  calculations  which will follow.   In Section
\protect\ref{sec:tree} we analyse  $q  \bar{q}$ annihilation into  two
gluons at tree level within the MYM model, and compare with the result
in   standard  YM.   In Section  \protect\ref{sec:loop}   we study the
one-loop contributions to $q \bar{q} \to q \bar{q}$ and show in detail
how the MYM model gives  rise to renormalisable and unitary $S$-matrix
elements.  
In Section \protect\ref{sec:2loop} we turn to the two-loop contribution
to $q \bar{q} \to q \bar{q}$, 
and  demonstrate the  emergence   of  logarithmically divergent
non-renormalisable  $S$-matrix     elements.        In      Section
\protect\ref{sec:Higgs} we  investigate  quantitatively the connection
of the MYM to field theories where the gauge  bosons acquire masses by
means  of the usual  Higgs mechanism  \cite{Higgs}.  In particular,  we
show how the presence of a Higgs boson 
 cancels the logarithmically 
non-renormalisable  contributions   found in  the   previous section.  
Throughout  Sections    \protect\ref{sec:tree}   --   
\protect\ref{sec:Higgs} we use  the pinch technique (PT) rearrangement
of  the $S$-matrix  \cite{PT,glmassJMC2}   in  order to   make several
cancellations manifest.  
We  hasten  to emphasise,  however,  that  the  PT  only  serves as  a
convenient  intermediate step, helping   to  expose the unitarity  and
renormalisation properties of  the $S$-matrix, but  none  of the final
results reported here depends  on the use of  this method.  In section
\protect\ref{sec:pomeron}   we  take  a   first  look  at  a  possible
phenomenological  application  of the   MYM model, namely, quark-quark
elastic scattering  in the  diffractive  region.  Finally, in  Section
\protect\ref{sec:conclusions}  we  summarise  our results and  discuss
possible future applications.

\section{The Massive Yang-Mills Model}

\protect\label{sec:mqcd}
In this section we first review briefly how local gauge-invariance and
massive gauge bosons can be reconciled in the MYM.   Next we show that
the MYM is physically equivalent to  a field theory  where the
gauge bosons have  been  endowed with  a  mass ``naively'',  i.e.   by
adding   a mass  term at   tree-level  without preserving gauge
invariance.
 
In order to introduce the MYM model \cite{Slavnov,Cornwall},  
let  us start from the standard YM action for the $SU(3)$ gauge group:
\begin{equation}
S_{YM} [A] = -\frac{1}{2} \int \ d^4 x \ {\rm Tr} ({\cal F}_{\mu \nu} {\cal F}^{\mu \nu}) 
\end{equation}
where ${\cal F}_{\mu \nu} (x)  = \partial_{\mu} A_{\nu} (x) - \partial_{\nu} A_{\mu} (x) + 
ig \ [A_{\mu} (x),A_{\nu} (x)]$ and $A_{\mu} (x)= A_{\mu}^{a} (x) 
T_{a}$, with $T_a$ the $SU(3)$ generators in the fundamental representation. 
For the purpose of the present discussion, matter fields can be ignored. 
Under a 
gauge transformation, parametrised by $U(x)$, $A_{\mu} 
\rightarrow A^U_{\mu}$ where 
\begin{equation}
A^U_{\mu} (x) \equiv U(x) \ A_{\mu} (x) \ U^{-1} (x) 
- \frac{i}{g} U (x) \partial_{\mu} U^{-1} (x).
\end{equation}
The requirement of gauge invariance for the action forbids a naive mass term for the gluon. 
However, by introducing $SU(3)$-valued fields, $V(x)$, one can define 
\begin{equation}
C_{\mu} (x) \equiv -\frac{i}{g} V (x)  \partial_{\mu} V^{-1} (x).
\label{eq:defc}
\end{equation}
Under a gauge transformation one postulates that 
$V  \rightarrow V^U =  U V$.  As  a 
consequence, $C_{\mu} (x)$ has the same gauge transformation properties 
as the gauge field $A_{\mu}$, i.e.
\begin{equation}
 C_{\mu}^U  =  U C_{\mu} U^{-1} - \frac{i}{g} U \partial_{\mu} U^{-1}.
\label{eq:c}
\end{equation} 
The quantity 
\begin{equation}
B_{\mu} [A,V] (x) \equiv A_{\mu} (x) - C_{\mu} (x) 
\label{eq:b}
\end{equation}
thus transforms as $B_{\mu}^U = U B_{\mu} U^{-1}$  under a simultaneous gauge 
transformation
of the  $A_{\mu}$  and $V$  fields, so one can add to the 
Yang-Mills action the following gauge-invariant term: 
\begin{equation}
{S}_M [A,V] =   M^{2} \int \ d^4 x \ {\rm Tr} \ B_{\mu} B^{\mu}. 
\label{eq:l_b}
\end{equation}
More explicitly, gauge invariance of the above quantity can be written as 
\begin{equation}
{S}_M [A^U,V^U] = {S}_M [A,V].
\end{equation}
Recalling (\ref{eq:defc}) and  (\ref{eq:b}),  
it is clear  that ${S}_M$ generates a mass term for 
the gluon field $A_{\mu}$, a kinetic term for the field $V$, and an interaction
term between the $A$ and $V$ fields. 

Finally, we can write down the 
gauge-invariant 
action functional for  the MYM theory:
\begin{equation}
S_{MYM} [A,V] \equiv S_{YM} [A] + S_M [A,V]. 
\end{equation}
We write now the path integral for such a theory.
 Gauge   invariance of $S_{MYM}$    implies  that a  
gauge-fixing  prescription   is needed   to    quantise the  theory.  The
Faddeev-Popov  procedure can be  carried out as  in standard YM theory, 
leading to 
\begin{equation}
{\cal Z} = \int \ {\cal D} V \ {\cal D} A \ e^{i S_{MYM} [A,V]} \ 
\Delta[A] \ \delta(g[A]).
\label{eq:fad}
\end{equation}  
 Here the gauge-fixing condition is $g[A]  = 0$ and $\Delta[A]$ is the 
corresponding Faddeev-Popov determinant . In  order to make the theory amenable
to a perturbative treatment one could rewrite ${S}_M [A,V]$ as a power
series in the coupling constant $g$. This is obtained by writing
\begin{equation}
V(x) = \exp(i g \ \theta^{a}(x) T_a), 
\label{angular}
\end{equation}
and inserting the power  expansion  for $V$ into  (\ref{eq:l_b}).  The
resulting expression  contains interaction vertices with an increasing
number of scalar $\theta$ fields and 
 zero or one  gauge field $A_{\mu}$. 
Then,    using standard techniques,  Feynman   rules   can be  derived
\cite{Slavnov}.
However,  as long as   one   is only  interested in   gauge-invariant
calculations,  a considerable simplification of  the Feynman rules can
be achieved.  To  see  this, let us  consider  the calculation  of the
vacuum expectation value of a generic gauge-invariant operator $O[A]$:
\begin{equation}
\langle O \rangle \equiv \frac{1}{{\cal Z}} \ 
\int \ {\cal D} V \ {\cal D} A \ e^{i S_{MYM} [A,V]} \
\Delta[A] \ \delta(g[A]) O[A].
\label{eq:vev}
\end{equation}
We perform a change of integration variable in the ${\cal D} A$ integral, 
i.e. we rewrite it in terms of a new field $A_{\mu}^{'}$,   
 defined through the following identity: 
\begin{equation}
A_{\mu} = V \ A_{\mu}^{'} V^{-1} - \frac{i}{g} V 
\partial_{\mu} V^{-1} \equiv A_{\mu}^{' \ V}. 
\end{equation}
In other words, $A_{\mu}$ and $  A_{\mu}^{'}$ are related by the gauge 
transformation generated by $V$. 
Thus, ${\cal D} A  = {\cal D} A^{'}$. Also,
 gauge invariance implies that
$S_{YM} [A] =  S_{YM}[A^{'}]$, 
$O[A]=O[A^{'}]$ and
 $\Delta[A]  =  \Delta[A^{'}]$. Strictly speaking, the last  equality 
holds only if one neglects the issue of Gribov copies.
This is correct for the purpose of a perturbative treatment.

The crucial observation is that because of (\ref{eq:defc}), 
(\ref{eq:c}) and (\ref{eq:b}) one has 
\begin{equation}
B_{\mu} [A,V] \equiv B_{\mu} [A^{' \ V},V]= V \ A_{\mu}^{'} V^{-1},
\end{equation}
hence we can write 
\begin{equation}
 S_M [A,V] =  M^2 \int \ d^4 x \ {\rm Tr} \ A_{\mu}^{'} A^{' \ \mu}. 
\end{equation}
The path integral (\ref{eq:vev}) can then be rewritten as
\begin{equation}
\langle O \rangle \equiv \frac{1}{{\cal Z}} \
\int \ {\cal D} V \ {\cal D} A^{'} \ {\rm e}^{i \left( S_{YM}[A^{'}] + 
M^2 \int d^4 x \ {\rm Tr} \ A_{\mu}^{'} A^{'\mu} \right)} \
\Delta[A^{'}] \ \delta(g[A^{' \ V}]) \ O[A^{'}].
\end{equation} 
Notice that in the above  expression all the dependence on the $V$ fields 
is carried by the $\delta$-function. 
The integration  on the $V$   fields yields a  factor
$1/\Delta[A^{'}]$,  which   cancels the  Faddeev-Popov determinant 
arising  from  the gauge-fixing procedure. The final path integral can be 
written as  
\begin{equation}
\langle O \rangle = \frac{1}{{\cal Z}} \
\int \ {\cal D}A \ {\rm e}^{i \left(S_{YM}[A] + 
M^2 \int d^4 x \ {\rm Tr} \ A_{\mu} A^{\mu} \right)} \ O[A].
\label{eq:mass}
\end{equation}
The above  manipulations show that, as  long as the operator
of interest is  gauge invariant in  the usual massless QCD  sense, the model
defined  by  (\ref{eq:fad})  is  equivalent to the simpler massive
vector theory defined by (\ref{eq:mass}). The latter is obviously
much easier to handle in perturbative calculations.
 
It is important to emphasise that the models are not equivalent at the
level of  (gauge dependent) Green's functions  of the gluon  field. In
particular, let us compare the tree-level  expressions
for the gluon propagator in the  two models.  From (\ref{eq:fad})
one obtains (in the Landau gauge)
\begin{equation} 
D_{\mu \nu}^{tree} (k) = {1 \over k^{2} - M^{2}} \ \left( g_{\mu \nu} - {k_{\mu} k_{\nu} \over k^{2}} 
\right), 
\label{eq:prop_fad}
\end{equation}
while (\ref{eq:mass}) yields
\begin{equation}
D_{\mu \nu}^{tree} (k) = {1 \over k^{2} - M^{2}} \ \left( g_{\mu \nu} - {k_{\mu} k_{\nu} \over M^{2}}
\right).
\label{eq:prop_mass}
\end{equation}
The  former expression 
 corresponds to a  gluon with two polarisation states, as
in the massless case, whilst the latter  has three polarisation states,
as expected
 for a massive vector boson.   Of course the  number of degrees  of  freedom in the two
models  has  to match.  In fact,  the third  polarisation state of  
(\ref{eq:prop_mass}) corresponds  to the massless scalar field $\theta
(x)$ which appears in the MYM model.
 
We have seen that gauge invariance can  be maintained in a theory of
massive gluons  without introducing additional particles into the
spectrum. However, as we  will discuss later,  and as  noted  by others
\cite{nonren1}, the resulting theory is no longer renormalisable.

\setcounter{equation}{0}
\section {Tree-level analysis}
\protect\label{sec:tree}

\begin{figure}[h] 
\centerline{\epsfig{file=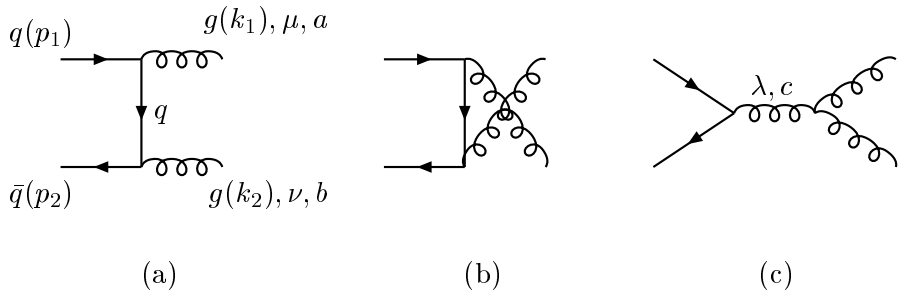,height=4cm,%
bbllx=71pt,bblly=510pt,bburx=405pt,bbury=622pt}}
\caption{Diagrams contributing to ${\cal T}^{ab}_{\mu \nu}$.}
\label{tree}
\end{figure}

In this section we will study in detail
the tree-level cross-section for quark--antiquark 
annihilation into two massive gluons, i.e.
$q\bar{q}\rightarrow gg$, within the framework of the MYM model.
The reason is three-fold: First we want to gain some familiarity with the
formalism, second we want to study the difference between the MYM and
standard QCD at the level of physical amplitudes, and third,
in conjunction with the results of the next section, we will check explicitly
that the MYM model produces unitary $S$-matrix elements.
 Throughout this section we use the methodology and notation first 
introduced in \cite{PP2}.

Let us consider the quantity ${\cal A}$,
\begin{equation}
{\cal A}
= \frac{1}{2} 
\int d(P.S. {}^2)\,  
\langle q\bar{q}|T|gg\rangle \langle gg|T|q\bar{q}\rangle^{\dagger}\, ,
\label{OTgg}
\end{equation}
where
\be
\int d(P.S.{}^2) = \frac{1}{(2\pi)^2}\, 
\int d^{4}k_1 \int d^{4}k_2\, \delta_+ (k^2_1-M^2)\delta_+ (k^2_2-M^2)
\delta^{(4)}(q-k_1-k_2)
\ee
is the phase space integral for two particles with equal mass $M$ in the 
final state,
with $\delta_+(k^2-M^2)\equiv \theta (k^0)\delta(k^2-M^2)$.
In (\ref{OTgg}), the factor 1/2 is statistical, arising from
the fact that the final on-shell gluons should be considered as identical
particles in the total rate. ${\cal A}$ is the contribution to the 
imaginary part of the amplitude for $q \bar{q} \to q \bar{q}$ which arises
from a gluon loop. We first focus on the tree-level amplitude
${\cal T}\equiv \langle q\bar{q}|T|gg\rangle$.
Diagrammatically, the amplitude ${\cal T}$ 
consists of two distinct parts: $t$ and $u$-channel graphs that contain an
internal quark propagator, ${{\cal T}_{t}}^{ab}_{\mu\nu}$, as shown in 
Fig.\ref{tree}(a,b) and an $s$-channel amplitude, 
${{\cal T}_{s}}^{ab}_{\mu\nu}$, as shown in Fig.\ref{tree}(c). 
The subscripts ``$s$'' and ``$t$'' refer to the
corresponding Mandelstam variables, i.e. $s=(p_1+p_2)^2=(k_1+k_2)^2$
and $t=(p_1-k_1)^2=(p_2-k_2)^2$. 

Let us first define the following quantities: 
\bea
V_{\rho}^{c}(p_1,p_2) &\equiv&g\bar{v}(p_2)\, T^c \gamma_{\rho}
\, u(p_1)\, ,\nonumber\\
{\cal R}_{\mu}^{ab}(p_1,p_2,q) &\equiv& gf^{abc}\,D_0(q) 
V_{\mu}^{c}(p_1,p_2)  \, ,
\label{VSG}
\eea
where
\be
D_0(q) \equiv \frac{1}{q^{2}-M^{2}}\, .
\ee
The amplitude is given by
\begin{equation}
{\cal T}^{ab}_{\mu\nu}={{\cal T}_{s}}^{ab}_{\mu\nu} +
{{\cal T}_{t}}^{ab}_{\mu\nu}\, ,
\label{DefT}
\end{equation}
with
\be
{{\cal T}_{s}}^{ab}_{\mu\nu}=
{\cal R}_{\lambda}^{ab}\Gamma^{\lambda}_{\mu\nu}(q,-k_1,-k_2) \, ,
\label{Ts}
\ee
where   
\begin{equation}
\Gamma_{\lambda\mu\nu}(q,-k_1,-k_2)\ =\ 
(q+k_1)_{\nu}g_{\lambda\mu}\, +\,
(k_2-k_1)_{\lambda}g_{\mu\nu}\, -\, (q+k_2)_{\mu}g_{\lambda\nu}\, ,
\label{3GV}
\end{equation}
is the usual three-gluon vertex and
\be
{{\cal T}_{t}}^{ab}_{\mu\nu} = -ig^2\bar{v}(p_2)\Big( 
\, T^b \gamma^{\nu}\, \frac{1}{\not\! p_1-\not\! k_1}
\, T^a \gamma^{\mu}\ +\
   T^a \gamma^{\mu}\, \frac{1}{\not\! p_1-\not\! k_2}
\, \gamma^{\nu} T^b \, \Big)u(p_1)\, .\qquad 
\label{Tt}
\ee
Notice that in (\ref{Ts}) only the `$g^{\mu\nu}$' part
of the tree-level massive gluon propagator appears,
since any longitudinal part vanishes due to current conservation 
when it hits the external on-shell quarks. The three-gluon vertex 
satisfies the fundamental Ward identity:
\bea
k_{1}^{\mu}\Gamma_{\lambda\mu\nu}(q,-k_1,-k_2) &=&
\big [d_0^{-1}(k_2)-d_0^{-1}(q)\big ]g_{\lambda\nu} + 
\big [q_{\lambda}q_{\nu}- k_{2\lambda}k_{2\nu}\big ]
\nonumber\\
&=&
\big [ D_0^{-1}(k_2)-D_0^{-1}(q) \big ] g_{\lambda\nu} + 
\big [q_{\lambda}q_{\nu}- k_{2\lambda}k_{2\nu}\big ]
\label{fundWI}
\eea
(and cyclic permutations) where $d_0^{-1}(q) \equiv q^2$.
The form of the Ward identity in the 
massive theory is therefore identical to that of the massless theory.

We then have
\begin{eqnarray}
{\cal A} &=& \frac{1}{2}
\, \int {\cal T}^{ab}_{\mu\nu}\, Q^{\mu\sigma}
(k_1)\, Q^{\nu\lambda}(k_2)\, {\cal T}^{ab\dagger}_{\sigma\lambda}d(P.S. {}^2)
\nonumber\\
&=& 
\frac{1}{2}\int\Big[ {{\cal T}_{s}}^{ab}_{\mu\nu}+
{{\cal T}_{t}}^{ab}_{\mu\nu}\Big]\, Q^{\mu\sigma}(k_1)\,
Q^{\nu\lambda}(k_2)\, \Big[ {{\cal T}_{s}}^{ab \dagger}_{\sigma\lambda}
+{{\cal T}_{t}}^{ab \dagger}_{\sigma\lambda}\Big] d(P.S. {}^2) ,
\label{MM}
\end{eqnarray}
where
\be
Q^{\mu\nu} (k)\ \equiv \ -\, g^{\mu\nu}\, +\, \frac{k^\mu k^\nu}{M^2}\, 
\ee
is the polarisation tensor of the massive gluon. 
On shell, i.e.  $k^2= M^2$, we have 
that $k^{\mu}Q_{\mu\nu}(k)=0$. 
This fact motivates the standard PT
decomposition of the three-gluon vertex \cite{CP}:
\begin{eqnarray}
\Gamma_{\lambda\mu\nu} (q,-k_1,-k_2) &=& 
\Gamma^F_{\lambda\mu\nu}(q,-k_1,-k_2)\ +\ 
\Gamma^P_{\lambda\mu\nu}(q,-k_1,-k_2)\,
\label{GFGP}
\end{eqnarray}
where
\begin{eqnarray}
\Gamma^F_{\lambda\mu\nu}(q,-k_1,-k_2) &=& (k_2-k_1)_{\lambda}g_{\mu\nu} 
+ 2q_{\nu}g_{\lambda\mu}-2q_{\mu}g_{\lambda\nu} \nonumber \\
\Gamma^P_{\lambda\mu\nu}(q,-k_1,-k_2) &=& k_{1\mu}g_{\lambda\nu}-
k_{2\nu}g_{\lambda\mu}.
\end{eqnarray}
The term $\Gamma^P_{\rho\mu\nu}$ vanishes when it hits the 
polarisation tensors, and  (\ref{MM}) becomes
\be
{\cal A} = 
\frac{1}{2}\int\Big[ {{\cal T}_{s}}^{F,ab}_{\mu\nu}+
{{\cal T}_{t}}^{ab}_{\mu\nu}\Big]\, Q^{\mu\sigma}(k_1)\,
Q^{\nu\lambda}(k_2)\, \Big[ {{\cal T}_{s}}^{F,ab \dagger}_{\sigma\lambda}
+{{\cal T}_{t}}^{ab \dagger}_{\sigma\lambda}\Big] d(P.S. {}^2) ,
\label{MMF}
\ee
where
\be
{{\cal T}_{s}}^{F,ab}_{\mu\nu} = 
{\cal R}_{\rho}^{ab}\Gamma_{\rho\mu\nu}^{F}.
\label{TF}
\ee
To evaluate further the expression on the RHS of 
 (\ref{MMF}) and establish its connection to massless QCD
we proceed to determine the action of the longitudinal momenta
coming from $Q^{\mu\sigma}(k_1)$
and $Q^{\nu\lambda}(k_2)$ on ${{\cal T}_{s}}^{F,ab}_{\mu\nu}$
and ${{\cal T}_{t}}^{ab}_{\mu\nu}$:
\bea
k_1^{\mu}{{\cal T}_{s}}^{F,ab}_{\mu\nu} & = &  
[(k_1-k_2)_{\lambda}k_{2\nu}-M^2g_{\lambda\nu}]{\cal R}_{\lambda}^{ab}
-D_0^{-1}(q){\cal R}_{\nu}^{ab}
\, ,
\label{w1}\\
k_2^\nu {{\cal T}_{s}}^{F,ab}_{\mu\nu}  & = &   
[(k_1-k_2)_{\lambda}k_{1\mu}+M^2g_{\lambda\mu}]{\cal R}_{\lambda}^{ab}
+D_0^{-1}(q){\cal R}_{\mu}^{ab}
\, ,\label{w2}\\
k_1^{\mu}{{\cal T}_{t}}^{ab}_{\mu\nu} & = & 
D_0^{-1}(q){\cal R}_{\nu}^{ab} \, ,
\label{w3}\\
k_2^{\nu}{{\cal T}_{t}}^{ab}_{\mu\nu} & = & -
D_0^{-1}(q){\cal R}_{\mu}^{ab} 
\label{w4}\, .
\eea
The terms proportional to $D_0^{-1}(q)$ cancel when forming the sum
$k^{\mu}_1 [{{\cal T}_s}^{F,ab}_{\mu \nu} + {{\cal T}_{t}}^{ ab}_{\mu\nu}]$, 
giving rise to
\begin{eqnarray}
k^{\mu}_1 [{{\cal T}_s}^{F,ab}_{\mu \nu} + {{\cal T}_{t}}^{ab}_{\mu\nu}]
&=&[(k_1-k_2)^{\lambda}k_{2\nu}-M^2
g^{\lambda}_{\nu}]{\cal R}_{\lambda}^{ab}
\, ,\nonumber\\
k^{\nu}_2 [{{\cal T}_s}
^{F,ab}_{\mu \nu} + {{\cal T}_{t}}^{ab}_{\mu\nu}]
&=& [(k_1-k_2)^{\lambda}k_{1\mu}+M^2
g^{\lambda}_{\mu}]{\cal R}_{\lambda}^{ab}
\, .
\label{BRSg1}
\end{eqnarray}
Such a cancellation is instrumental for the good high-energy 
behaviour of the resulting amplitudes.
Using the longitudinal momenta inside the polarisation 
tensors to trigger the identities listed above, 
we can decompose ${\cal A}$ into three parts:
\be
{\cal A}={\cal A}_{1}+{\cal A}_{2}+{\cal A}_{3}
\ee
where
\begin{eqnarray}
{\cal A}_{1} &=& 
\frac{1}{2}\, 
\int\Big[{\cal T}^{F}_{s}{{\cal T}^{F}_{s}}^{\dagger}
-R_{\mu} \big\{~\frac{7}{4} (k_1-k_2)^{\mu}(k_1-k_2)^{\nu}   
+2 M^2 g^{\mu\nu}\big\}~R_{\nu}^{\dagger} \Big] d(P.S. {}^2), 
\label{A1}\\
{\cal A}_{2} &=& \frac{1}{2}\int
( {\cal T}_{t} {{\cal T}^{F}_{s}}^{\dagger}
+{\cal T}^{F}_{s}{\cal T}_{t}^{\dagger} ) d(P.S. {}^2),
\label{A2} \\
{\cal A}_{3} &=& \frac{1}{2}\int
{\cal T}_{t}{\cal T}_{t}^{\dagger} d(P.S. {}^2).
\label{A3}
\end{eqnarray}
${\cal A}_1$   contains  the  purely  propagator-like   (self-energy)
contributions, ${\cal A}_2$ contains the vertex-like contributions and
${\cal A}_3$  contains  the box-like contributions.   We  see that all
terms  proportional to  $M^{-2}$ or  $M^{-4}$   have disappeared. {\it
  Therefore, at this point, it is clear that at the one-loop level the
  MYM model gives  rise to a renormalisable  $S$-matrix for $q \bar{q}
  \to q \bar{q}$, provided  that we assume unitarity and analyticity}
(i.e. dispersion relations).  In the next  section we shall check this
conclusion by an explicit one-loop calculation.

We now focus on the propagator-like part, ${\cal A}_{1}$.
Current conservation allows us to make the replacement
\be
\Gamma^{F}_{\rho\mu\nu}\Gamma^{F,\mu\nu}_{\lambda} \ \to \ 
8q^2g_{\rho\lambda} + 4{(k_1-k_2)}_{\rho}{(k_1-k_2)}_{\lambda}\, .
\label{FF}
\ee
Then (\ref{A1}) becomes
\be
{\cal A}_{1} =\ g^2 \, c_A \, D_0^2(q) \,
V_{\mu}^{c}\Biggl\{\int \left[ (4q^{2}-M^{2})g^{\mu\nu}+
\frac{9}{8}(k_1-k_2)^{\mu}(k_1-k_2)^{\nu} \right]  d(P.S. {}^2)
\Biggr\}V_{\nu}^{c}\, ,
\label{FM2}
\ee
where $c_A$ is the Casimir eigenvalue in the adjoint representation. 
The final step is to use the following results for the 
phase space integrals:
\bea
&& \int d(P.S.{}^2) =
\frac{1}{8\pi}\, \theta(q^0)\theta (q^2-4M^2)\, 
\Delta(q^2)\, ,\nonumber\\
&&\int d(P.S.{}^2) {(k_1-k_2)}_{\mu}{(k_1-k_2)}_{\nu} =
- \frac{1}{24\pi}\theta(q^0)\theta (q^2-4M^2 )\, 
q^{2}\Delta^{3}(q^2)g_{\mu\nu}\, ,
\label{LIPS12}
\eea
where 
\begin{equation}
\Delta(q^{2}) \equiv \sqrt{1-\frac{4M^{2}}{q^2}}.
\label{delta} 
\end{equation}
We obtain 
\bea
{\cal A}_{1}\ &=&\ \, D_0^2(q)  \,
V_{\mu}^{c}\Biggl\{
 \frac{\alpha_s}{2}\, 
c_A q^2 \Delta(q^2) 
\left(\frac{29}{8}+ \frac{1}{2}\frac{M^{2}}{q^{2}}\right)g^{\mu\nu}
\Biggr\}V_{\nu}^{c},
 \,\nonumber\\
&=& \ \, D_0^2(q) \,
V_{\mu}^{c}\Biggl\{
\frac{\alpha_s}{2}\,c_A q^2 \Delta(q^2)
\Bigg [\left(\frac{11}{3}-\frac{1}{24}\right)+ 
\frac{1}{2}\frac{M^{2}}{q^{2}}\Bigg ]
g^{\mu\nu}\Biggr\}V_{\nu}^{c}
\label{IMMYM}
\eea
with $\alpha_s=g^2/(4\pi)$. The reason why we write
the coefficient $\frac{29}{8}$ as the deviation from 
$\frac{11}{3}$ on the second line 
of Eq. (\ref{IMMYM}) will become clear in what follows.

It is instructive to repeat the same calculation for the case of massless
QCD, in order to examine the physical difference between the two theories
at tree-level \cite{PP2}.
The crucial modification, in the case of QCD, 
is that in  (\ref{MM}) the polarisation
tensors $Q_{\mu\nu}$, corresponding to the massive gluons,
are replaced by the polarisation tensors
$P^{\mu\nu}(k,\eta )$, given by
\begin{equation}
P_{\mu\nu}(k,\eta )\ =\ -\, g_{\mu\nu} + \, \frac{\eta_{\mu}k_{\nu}
+\eta_{\nu}k_{\mu} }{\eta k} - 
\eta^2 \frac{k_{\mu}k_{\nu}}{{(\eta k)}^2}\, ,
\label{PhotPol}
\end{equation}
which are appropriate for massless spin-1 gauge bosons.
As before we have that, for massless on-shell gluons,  $k^{\mu}P_{\mu\nu}=0$. 
All other expressions can be obtained directly from
the MYM expressions simply by setting $M^{2}=0$.
In particular, both the derivation and the final form of the 
Ward identities of (\ref{BRSg1})
are identical  \cite{PP2,ChengLi}

The QCD expression corresponding to (\ref{FM2}) 
is given by \cite{PP2}
\begin{equation}
{\cal A}_{1}^{QCD}\ =\ g^2 \, c_A  \, d_0^2(q) \,
V^{c}_{\mu}  \, \Biggl\{ \int \big[ 4q^2 g^{\mu\nu} \, +\,
{(k_1-k_2)}^{\mu}{(k_1-k_2)}^{\nu}\big] d(P.S.{}^2)\Biggr\}
\, V^{c}_{\nu}
\label{FM2QCD}
\end{equation} 
and, after carrying out the phase space integration for the two
final (massless) gluons, we obtain the QCD analogue of  (\ref{IMMYM}):
\begin{equation}
{\cal A}_{1}^{QCD} \ =
V^{c}_{\mu}\, d_0^2(q) \Biggl\{  \frac{\alpha_s}{2}\, 
\bigg(\frac{11}{3}\bigg)\,c_A q^2 g^{\mu\nu}\Biggr\} \, V^{c}_{\nu}\, ,
\label{IMQCD}
\end{equation}
Notice that the factor $\frac{11}{3}$ in (\ref{IMQCD}) is
the characteristic coefficient of the one-loop
$\beta$ function of quarkless QCD.

Obviously, if we set $M^{2}=0$ in  (\ref{FM2})
and  (\ref{IMMYM})
we do not recover the massless
QCD result, i.e. ${\cal A}_{1}(M^2=0) \neq {\cal A}_{1}^{QCD}$.
In that limit the two answers differ by the amount
$\frac{1}{24}$; 
this descrepancy heralds the  
difference in 
the leading logarithmic
behaviour of the two theories, which we will
establish in the next section.
On the other hand, it is clear that 
${\cal A}_{i}(M^2=0)={\cal A}_{i}^{QCD}$ for $i=2,3$. 
Evidently, even though the two theories satisfy the same type of 
tree-level Ward identities,
the fact that we have to use different polarisation tensors for
massive and massless gluons 
gives rise to different $S$-matrix elements, and this difference 
persists even in the limit $M\rightarrow 0$. 
As explained by Slavnov \cite{Slavnov}, the physical reason why
the limit $M\rightarrow 0$ of MYM does not recover massless Yang-Mills is
that one cannot continuously go from three polarisation states to two. 
It is interesting to notice that after the PT rearrangement
the discrepancy between the two theories as $M\rightarrow 0$
has been isolated in the universal, process-independent, 
propagator-like piece, ${\cal A}_1$.

\setcounter{equation}{0}
\section{One-loop analysis} 
\protect\label{sec:loop}

In   this   section we    turn   to  the   issue   of   unitarity  and
renormalisability at one-loop. To begin with, we show using a one-loop
calculation that  all unphysical poles  introduced by the gauge-fixing
choice cancel  in an   $S$-matrix element.    This cancellation is   a
necessary condition   for  proving the   unitarity  of  the  resulting
expressions;   indeed,  if expressions   containing   mixed poles  had
survived, they would give  rise to unphysical thresholds.  Next,
by comparing  the results of this section  with  those of the previous
one, we will be able to establish explicitly the validity of the
generalised optical theorem to lowest
order  and hence  have an explicit  demonstration  of unitarity at one
loop. Finally,  we show that  the  resulting expressions can  be  made
finite by the   usual     mass and wave-function    renormalisation.   
Throughout  this section we employ   the PT, which makes cancellations
particularly easy to track down.

We study the one-loop amplitude, 
${\cal M}=\langle
q\bar{q}|T|q\bar{q}\rangle, $ 
for the process $q\bar{q}\rightarrow q\bar{q}$, using the Feynman rules 
derived by Slavnov \cite{Slavnov}: the massive gluon propagator in 
the Landau gauge 
\footnote{Slavnov's choice of the Landau gauge was motivated by the fact
that it leads to a reduction in the number of interaction vertices.
Of course, for computations of $S$-matrix elements any other
choice of the gauge fixing parameter will lead to the same final answer}
is given by,
\bea
D_{\mu \nu} (k) &=& 
{1 \over k^{2} - M^{2}} \ \left( g_{\mu \nu} - {k_{\mu} k_{\nu} \over k^{2}} 
\right),\nonumber\\ 
&\equiv& D_0(k) \, t_{\mu \nu},
\label{slavprop}
\eea 
the ghosts are massless and only appear inside closed loops (with
a statistical factor $-1/2$), and the three- and 
four-gluon vertices are those known from massless QCD. Note that
we do not include quark loops since they are trivially
related to the equivalent QCD diagrams and, as such, need not be considered
when investigating the new features of the MYM model.
We will show explicitly that all unphysical poles (i.e. massless poles in
the Landau gauge) induced by the 
longitudinal part of $D_{\mu \nu}$ and by the massless ghosts
 vanish in the one-loop amplitude ${\cal M}$. 
Moreover, all contributions containing unphysical poles are
propagator-like, in the sense defined by the PT re-arrangement
of the amplitude \cite{PT,glmassJMC2}.

First we define 
\bea
I(q,k)&\equiv&\ \frac{1}{(k^2-M^2)[(k+q)^2-M^2]} \, ,\nonumber\\
J(k)&\equiv& \frac{1}{k^2-M^2} \, ,
\eea
and the auxiliary expressions containing mixed poles:
\bea
I_0(q,k)&\equiv&\ \frac{1}{k^2(k+q)^2}\ , \nonumber\\ 
I_1(q,k)&\equiv& \ \frac{1}{k^2(k^2-M^2)[(k+q)^2-M^2]}\ ,\nonumber\\
I_2(q,k)&\equiv& \ \frac{1}{k^2(k+q)^2(k^2-M^2)[(k+q)^2-M^2]}\ ,\nonumber\\
J_1(k)&\equiv& \frac{1}{k^2(k^2-M^2)}\ , 
\eea
which appear in intermediate steps but  vanish in the
final answer. In addition, we define     
\be
U_{\mu\nu}^{-1}(k) \equiv  D_0^{-1}(k)g_{\mu\nu}-k_{\mu}k_{\nu}. 
\ee 

\begin{figure}[t] 
\centerline{\epsfig{file=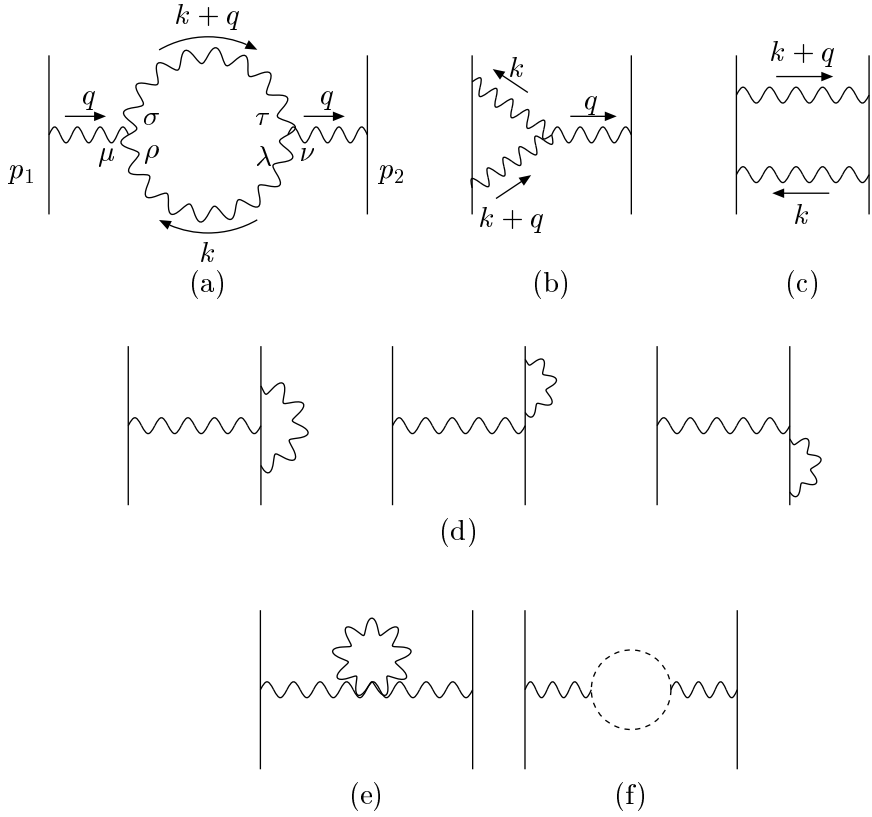,height=11cm,%
bbllx=74pt,bblly=379pt,bburx=407pt,bbury=692pt}}
\caption{One loop diagrams.}
\label{loop}
\end{figure}

We consider the diagrams of Fig.2 individually.
The expressions for all non-propagator-like contributions are the same  
as the corresponding contributions of the
regular QCD graphs in the Feynman-'t Hooft gauge, with the only
difference that the internal  gluon  propagators are  $D_0$ rather than
$d_{0}$.  These results emerge at the end of a
gauge-invariant calculation and are not linked to any particular
gauge choice. Consequently we turn our attention to the propagator-like
contributions.

For each diagram of Fig.2 we write the associated amplitude as a sum of 
propagator-like ($P$) and non-propagator-like ($NP$) pieces:
\begin{equation}
{\cal M}^{(i)} = {\cal M}^{(i)}_P + {\cal M}^{(i)}_{NP},
\end{equation}
where $i$ labels which diagram is being considered. Of course
${\cal M}^{(a)}_{NP} = {\cal M}^{(e)}_{NP} = {\cal M}^{(f)}_{NP} = 0$.
For the propagator-like piece it is convenient to write
\begin{equation}
{\cal M}^{(i)}_P = g^2 \ c_A \ D_0^2(q) \ V^{\mu}_c V^{\nu}_c
\int \frac{d^n k}{i(2 \pi)^n} \ \Pi^{(i)}_{\mu \nu}.
\end{equation} 

For graph ($a$) one finds that
\begin{equation}
\Pi^{(a)}_{\mu \nu}  = \Pi^{(a)}_{0,\mu \nu} + \Pi^{(a)}_{1,\mu \nu}
\end{equation} 
where 
\begin{equation}
\Pi^{(a)}_{0,\mu \nu} = -\frac{1}{2} I(q,k) \
\Gamma_{\mu\rho\sigma}(q,k,-k-q)
\Gamma_{\nu}^{\rho\sigma}(q,k,-k-q) 
\end{equation}
and
\begin{eqnarray}
\Pi^{(a)}_{1,\mu \nu} &=& -I_1(q,k) \Bigg [ D_{0}^{-2}(q) g_{\mu\nu} - 
2 D_{0}^{-1}(q) U_{\mu\nu}^{-1}(k+q)
+U_{\mu\rho}^{-1}(k+q) U^{\rho -1}_{\nu}(k+q)\Bigg ] \nonumber \\ 
&+& \ \frac{1}{2} I_2(q,k) \Bigg[D_{0}^{-2}(q) + 2 M^2 D_{0}^{-1}(q) + 
M^4 \Bigg] k_{\mu}k_{\nu}.
\label{figa}
\end{eqnarray}
$\Pi^{(a)}_0$ is the part of $(a)$ which arises due to the 
$g^{\rho\lambda} g^{\sigma\tau}$ part of $t^{\rho\lambda}t^{\sigma\tau}$
and contains only the physical (massive) poles. The unwanted mixed poles
live in $\Pi^{(a)}_1$. We note that the term in parenthesis accompanying
the $I_2$ factor of (\ref{figa}) is equal to $q^4$.  We choose not to
simplify this expression since the terms proportional to
inverse powers of $D_0$ are going to cancel against similar
contributions from vertex and box graphs; retaining them explicitly will 
make the mechanism of the cancellation more transparent. 

We next turn to the  vertex graph of Fig.2(b) (and its mirror image).  
We write
\begin{equation}
\Pi^{(b)}_{\mu \nu}  = \Pi^{(b)}_{0,\mu \nu} + \Pi^{(b)}_{1,\mu \nu}.
\end{equation} 
The contribution which arises from the `$g^{\alpha \beta} g^{\delta \gamma}$' 
term in the product of the two gluon polarisation tensors
is equal to the usual QCD vertex graph in the Feynman-'t Hooft gauge with 
massive, instead of massless, internal gluon propagators. This term can still
give a propagator-like contribution due to the pinching of the fermion
propagator triggered by the three-gluon vertex \cite{PT}. 
This contribution is  
\begin{equation}
\Pi^{(b)}_{0,\mu \nu} = 2 I(q,k) \ D_0(q)^{-1} \ g_{\mu \nu}.
\end{equation}
The $\Pi^{(b)}_1$ term contains the remaining parts of the polarisation
tensor product and the pinching of the quark propagator is triggered by 
the momenta therein:
\begin{eqnarray}
\Pi^{(b)}_{1,\mu \nu} &=& 2 D_{0}^{-1}(q) I_1(q,k) \Bigg[ 
D_0^{-1}(q) g_{\mu\nu} - U_{\mu\nu}^{-1}(k+q)\Bigg]
\nonumber\\ & &  -  D_{0}^{-1}(q) I_2(q,k) \Bigg[ D_0^{-1}(q) + M^2
\Bigg] k_{\mu}k_{\nu}.
\label{figb}
\end{eqnarray}

For the box graph of Fig.2(c) (along with the crossed box):
\begin{eqnarray}
\Pi^{(c)}_{\mu \nu} &=& - D_0^{-2}(q) \ I_1(q,k) \ g_{\mu\nu}
+ \frac{1}{2} D_0^{-2}(q) \ I_2(q,k) \ k_{\mu} k_{\nu},
\label{figc}
\end{eqnarray}
and for the remaining graphs: 
\begin{eqnarray}
\Pi^{(d)}_{\mu \nu} &=& D_{0}^{-1}(q) \ J_1(k) g_{\mu\nu} \, , 
\\
\Pi^{(e)}_{\mu \nu} &=&  k^2 J_1(k) \ t_{\mu\nu}(k) \, , 
\\
\Pi^{(f)}_{\mu \nu} &=& -\frac{1}{2} I_0(q,k) \ k_{\mu}k_{\nu} \, .
\end{eqnarray}

Notice that, at this point, all terms containing massless propagators
are multiplied by inverse powers of $D_0(q)$. If we now add these 
 contributions, all terms proportional to inverse powers of 
$D_0(q)$, and therefore all terms containing
massless poles, cancel against each other. In order for this cancellation 
to go through it is crucial that the ghost diagram has a statistical 
factor of $(-1/2)$, rather than the $-1$ of massless QCD.
It is also interesting to observe that the aforementioned cancellations 
take place algebraically before 
any of the integrations over the virtual momenta
$k$ are carried out. In particular, we have not resorted to the
use of dimensional regularisation results such as 
$\int \frac{d^nk}{k^2}=0$ or  $\int \frac{d^nk}{k^4}=0$.

Our final result for the propagator-like part of the one-loop
amplitude is thus
\begin{equation}
\Pi_{\mu \nu} =  \Pi^{(a)}_{0,\mu \nu} + 
I(q,k) \bigg( 2 D_0^{-1}(q) g_{\mu\nu}
- \frac{1}{2} k_{\mu}k_{\nu}  \bigg).
\label{full1}
\end{equation}
Notice that the last term in the above equation could be interpreted as
a contribution from massive ghosts. This term has emerged naturally 
from our calculation, even though we started out with massless ghosts.

As we have already mentioned, in the limit $M\rightarrow 0$, 
the box-like and vertex-like parts of the MYM $S$-matrix element
exactly reproduce their massless QCD counterparts. 
The propagator-like piece is, as expected from the work of the previous
section, different. Defining the effective gluon self-energy,
$\widehat{\Pi}$, via
\begin{equation}
{\cal M}_P = D_0^2(q) V^{\mu}_c \widehat{\Pi}(q^2) 
V_{\mu c} 
\end{equation}
a straightforward calculation yields
\begin{eqnarray}
\widehat{\Pi}(q^2) &=& g^2 c_A \Biggl\{
\Bigg[ \frac{29}{8}q^2+\frac{1}{2}M^2 + \frac{\epsilon}{24}
(q^2-4M^2) \Bigg] \int \frac{d^n k}{i (2 \pi)^n} I(q,k) \nonumber \\
& & - \Bigg (\frac{5}{4} - \frac{11\epsilon}{12}\Bigg)
\int \frac{d^n k}{i (2 \pi)^n} J(k) \Biggr\} 
\label{M1fin}
\end{eqnarray}
where $\epsilon=4-n$. 
Setting $M^2=0$,
and using
$\int \frac{d^n k}{k^2}=0$, 
the above expression reduces to
\be
\left. \widehat{\Pi}(q^2)\right|_{M=0}
=  g^2 c_A \ q^2 \left( \frac{29}{8} + \frac{\epsilon}{24} \right) 
\int \frac{d^n k}{i(2 \pi)^n} I_0(q,k).
\ee

The corresponding 
gauge-invariant effective self-energy for
massless QCD, first given in \cite{CP}, reads:
\be
\widehat{\Pi}^{QCD}(q^2) = g^2 c_A
q^2 \left( \frac{11}{3} + \frac{\epsilon}{6} \right)
\int \frac{d^n k}{i(2 \pi)^n} I_0(q,k).
\ee
Notice that the above result 
can be obtained from (\ref{full1}) by taking the
$M^2\to 0$ limit and, at the same time, changing by hand the
coefficient in front of the ghost term from (-1/2) to 
(-1).

To establish contact with the previous section, we need to compute
the imaginary part of ${\cal M}_P$. Using
\begin{eqnarray}
\Im {\rm m}\Big[\int \frac{d^nk}{i(2\pi)^n}\, I(q,k)\Big] &=&
-\frac{1}{16\pi^2}\, \Im {\rm m}\Big\{ \, \int^1_{0} dx
\ln[M^2-q^2x(1-x)]\Big\}
\nonumber\\
&=& \frac{\theta ( q^2-4M^2 )}{16\pi}\, 
\Delta(q^2)\nonumber\\
&=& \frac{1}{2}\int d(P.S.{}^2)\, ,\qquad
\label{imI}
\end{eqnarray} 
it is straightforward to check that unitarity holds,
i.e. 
\be
2 \Im {\rm m} {\cal M}_P = {\cal A}_1.
\ee
Similarly, one can demonstrate the unitarity of the 
vertex- and box-like contributions.

We next proceed to 
renormalise the expression for 
$\widehat{\Pi}(q^2)$; we carry out the two subractions 
(corresponding to mass and wave-function renormalisation)
at $q^2=M^2$ (``on-shell'' scheme
\footnote{
Any other subtraction point $\mu^2$ would work equally well.}) i.e.
\be
\widehat{\Pi}_{R}(q^2) = \widehat{\Pi}(q^2) - \widehat{\Pi}(M^2) -
 (q^2-M^2) \left. \frac{\partial\widehat{\Pi}(q^2)}
{\partial q^{2}}\right|_{q^2=M^{2}}
\label{PR1}
\ee
and so the renormalised self-energy $\widehat{\Pi}_{R}(q^2)$ becomes 
\be
\widehat{\Pi}_{R}(q^2) =
\frac{\alpha_s c_A}{4\pi} \Biggl\{
\Bigg(\frac{29}{8}q^2+\frac{1}{2}M^2 \Bigg)\Bigg ( L(q^2)-L(M^2) \Bigg )
-\frac{11}{8}(q^2-M^2)\Bigg(3-2L(M^2)\Bigg )\Biggl\}
\label{PR2}
\ee
where
\be
L(q^2)=\Delta(q^2)\ln 
\biggl(\frac{\Delta(q^2)+1}{\Delta(q^2)-1}\biggr)
\ee
and $\Delta(q^2)$ was defined in (\ref{delta})~\cite{PRW}.
Note that only the self-energy contribution $\widehat{\Pi}(q^2)$ needs
to be renormalised; indeed, after  the PT rearrangement the  resulting
expressions  for the  vertices  (and boxes)  are  ultra-violet finite,
exactly as  happens in  normal QCD  \cite{PT}. As  a result, the gluon
wave-function  renormalisation  constant $Z_A$  and the gauge coupling
renormalisation constant $Z_g$ are   related by the QED-like  relation
$Z_{A}=Z_{g}^{1/2}$ \cite{PT,PP2,CP}.

In the limit $q^2\gg M^2$, the leading (logarithmic) 
contribution to $\widehat{\Pi}_{R}(q^2)$ is given by
\be
\widehat{\Pi}_{R}(q^2)=  \frac{\alpha_s c_A}{4\pi}
 \left( \frac{29}{8} \right) q^2 \ln(q^2/\mu^2) + ...
\label{PR2Limit}
\ee
where the ellipsis denotes subleading contributions and 
$\mu$ is an arbitrary reference momentum.
Instead, the corresponding limit for
QCD is given by 
\be
\widehat{\Pi}_{R}^{QCD}(q^2)=
\frac{\alpha_s c_A}{4\pi} \left
( \frac{11}{3} \right) q^2 \ln(q^2/\mu^2) + ... 
\ee

 It is also interesting to compare 
the qualitative features of 
the MYM self-energy $\widehat{\Pi}$ with
Cornwall's massive propagator \cite{glmassJMC2}
which has been used successefully for fitting
data \cite{diffraction2}; it 
has the form 
\footnote{The functional form
 for $d_C^{-1}(q^2)$ given in (\ref{givenbelow}) 
represents an excellent,
physically motivated fit to the numerical 
solution of a Schwinger-Dyson
equation for the gauge-independent QCD gluon self-energy.} 
(for Euclidean $q^2$)
\be
\label{CornProp}
d_C^{-1}(q^2)= [q^2+m^2(q^2)]bg^2\ln\Bigg[
\frac{q^2+4m^2(q^2)}{\Lambda^2}\Bigg]
\label{givenbelow} 
\ee
with
\be
\label{CornMass}
m^2(q^2)=m^2\Bigg[\frac{\ln[\frac{q^2+4m^2}{\Lambda^2}]}
{\ln(\frac{4m^2}{\Lambda^2})}\Bigg]^{-12/11} ~~,
\ee
where $\Lambda$ is the QCD mass.
Both $ d_C^{-1}(q^2)$
and $\widehat{\Pi}(q^2)$
display the  
correct threshold behaviour (i.e. they turn imaginary
for $-q^2=4m^2$). In addition (and in contrast to
$\widehat{\Pi}(q^2)$)  $ d_C^{-1}(q^2)$ has    
the correct asymptotic limit
for $q^2 \gg \Lambda^2$, since the coefficient multiplying
the leading logarithm is $11/3$ (instead of $29/8$ in the
case of $\widehat{\Pi}(q^2)$) ,
thus capturing the one-loop QCD running coupling.
Notice also the non-trivial dependence of the mass $m(q^2)$ 
on the momentum. 

Finally, it is straightforward to check that if one inserts the expression
for $\Im {\rm m} \widehat{\Pi}(q^2)$ obtained from the tree-level
calculation of the previous section into a twice-subtracted dispersion
relation then one 
 obtains the real part of the right hand side of 
 (\ref{PR2}), i.e.
\be
\label{RDRW}
\Re {\rm e}\widehat{\Pi}_{R}(s) =
\frac{(s-M^2)^2}{\pi}\int\limits^\infty_{4M^2}
\, \frac{ ds'\, \Im {\rm m} \widehat{\Pi}(s')}{(s'-M^2)^2 (s'-s)}\, .
\ee

We end this section by commenting on how the naive massive model gives 
precisely the same result for the one-loop $S$-matrix element in question. 
We know that this must be the case, given the work of 
Section \protect\ref{sec:mqcd}. 
The equivalence also follows from the tree-level arguments of the 
previous section. To see this, note that in computing the imaginary part 
of the one-loop amplitude we needed the $q \bar{q} \to gg$ amplitude only 
(i.e. in the unitarity equation the sum is over physical states and so ghost 
states do not contribute). In addition, the internal gluon propagators 
couple to conserved currents and so are the same in both MYM and naive 
gluon calculations. Under the assumption of analyticity, it follows that 
the two approaches give the same one-loop amplitude.
To see how things go working explicitly with the full one-loop
amplitude one needs to repeat the calculations of this section.
The only differences between the $S$-matrix element of the MYM compared to 
the naive model are the replacement of the bare gluon propagator of 
 (\ref{slavprop}) with the unitary gauge propagator:
\be
U_{\mu\nu}(q)= 
\Bigg ( g_{\mu\nu}- \frac{q_{\mu}q_{\nu}}{M^{2}}\Bigg )\frac{1}{q^2-M^2}
\, ,
\label{uniprop}
\ee
and the fact that the naive model does not have any ghosts. 
The actual calculation is straightforward, given the results presented 
above. One needs to replace the massless poles appearing in the auxiliary 
integrals $I_{1}$, $I_{2}$, and $J_{1}$, stemming from the longitudinal 
part of the gluon propagator, by $M^2$.  The algebraic cancellations go 
through in exactly the same way as before with $\frac{1}{k^2} 
\rightarrow  \frac{1}{M^2}$ and $\frac{1}{k^2(k+q)^2} \rightarrow  
\frac{1}{M^4}$.

\setcounter{equation}{0}
\section {Two-loop analysis}
\protect\label{sec:2loop}

Now we turn to the two-loop calculation. We will show that in this case
renormalisability breaks down, and that the non-renormalisable terms are
propagator-like and depend logarithmically on the cutoff.
We will work again directly with the
$S$-matrix element for the process $q\bar{q}\to q\bar{q}$. The 
calculations will be carried out using
the Feynman rules for the naive massive gluon
model since we know that, at the $S$-matrix level, it is equivalent to the 
MYM. 

Consider the tree-level amplitude of Section \ref{sec:tree},
${\cal T}_{0\mu\nu}^{ab}$ (note that we have changed notation by
adding the subscript `0' to denote a tree-level amplitude). 
It satisfies the following BRST identities \cite{ChengLi}:
\begin{eqnarray}
k^{\mu}_1{\cal T}_{0\mu\nu}^{ab} &=& k_{2\nu}{\cal S}_{0}^{ab}\, ,
\nonumber\\
k^{\nu}_2{\cal T}_{0\mu\nu}^{ab} &=& k_{1\mu}{\cal S}_{0}^{ab}\, ,  
\nonumber\\
k^{\mu}_1k^{\nu}_2{\cal T}_{0\mu\nu}^{ab} &=& M^2 {\cal S}_{0}^{ab}\, ,
\label{BRSg}
\end{eqnarray}
where 
\be
{\cal S}_0^{ab}=gf^{abc}\, \frac{k^\sigma_1}{q^2}\, 
V_{\sigma}^{c}= gf^{abc}\,\frac{k^\sigma_2}{q^2}\, V_{\sigma}^{c}.
\ee
Using (\ref{BRSg}) one finds that the imaginary part of the
amplitude, ${\cal A}$, can be written:
\bea
{\cal A} &=& \frac{1}{2}\int\,  {\cal T}_{0\mu\nu}\, 
Q^{\mu\rho}(k_1)\, Q^{\nu\sigma}(k_2)\, 
{\cal T}_{0\rho\sigma}^{\dagger} d(P.S. {}^2)
\nonumber\\  
&=& \frac{1}{2}\, \int\Big( {\cal T}_{0}^{\mu\nu} 
{\cal T}_{0\mu\nu}^{\dagger}\ -\ 
{\cal S}_{0}{\cal S}_{0}^{\dagger}\Big) d(P.S. {}^2) \ ,
\label{BRSMYM}
\eea
whereas for normal Yang-Mills:
\bea
{\cal A}_{QCD} &=& \frac{1}{2}\,  \int {\cal T}_{0\mu\nu}\, 
P^{\mu\rho}(k_1,\eta)\, P^{\nu\sigma}(k_2,\eta)\, 
{\cal T}_{0\rho\sigma}^{\dagger} d(P.S. {}^2)
\nonumber\\
&=& \frac{1}{2}\, \int
\Big( {\cal T}_{0}^{\mu\nu} {\cal T}_{0\mu\nu}^{\dagger}\ -\ 
2 {\cal S}_{0}{\cal S}^{\dagger}_{0}\Big) d(P.S. {}^2).
\label{BRSQCD}
\eea
Notice that, despite the different factors accompanying the 
${\cal S}_0{\cal S}^{\dagger}_{0}$ terms in (\ref{BRSMYM}) and 
(\ref{BRSQCD}), both expressions give rise to renormalisable real parts, i.e.
the real part can be obtained by means of a twice-subtracted 
dispersion relation. Renormalisability is manifest since the tree-level
amplitudes contain no dangerous terms (such terms vanish by current
conservation) and the contraction via the polarisation tensors does not
induce any non-renormalisable terms, as a consequence of (\ref{BRSg}).

\begin{figure}[h] 
\centerline{\epsfig{file=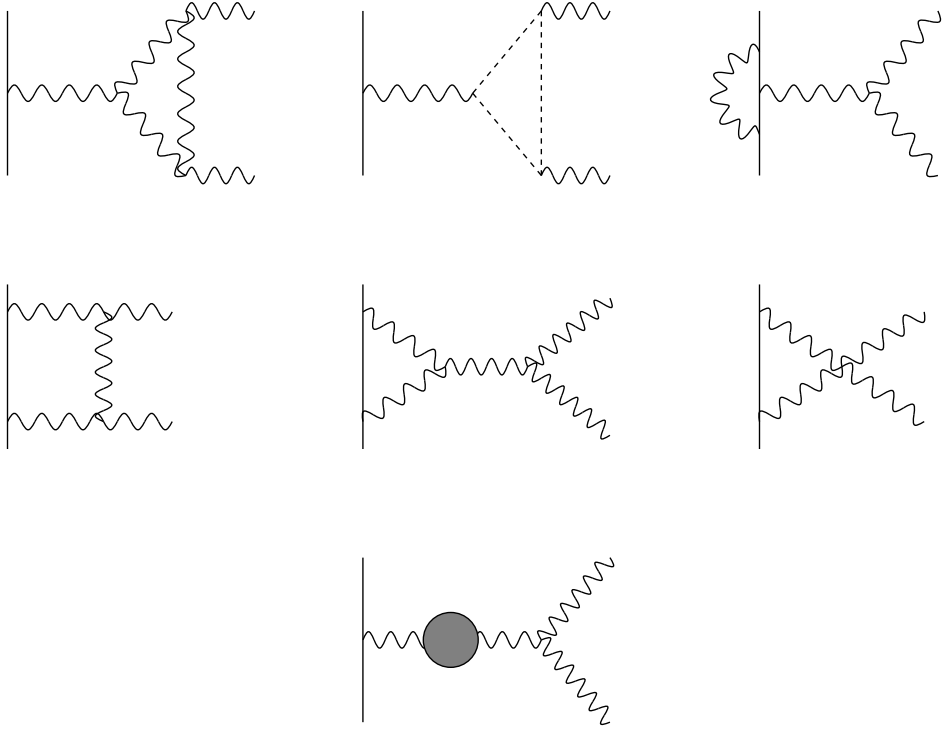,height=10cm,
bbllx=90pt,bblly=420pt,bburx=440pt,bbury=695pt}}
\caption{Diagrams contributing to ${\cal T}^{ab}_{1\mu \nu}$. The blob
refers to the same corrections to the gluon propagator as in 
Fig.\ref{loop}(a,e,f).}
\label{T1}
\end{figure}

\begin{figure}[h] 
\centerline{\epsfig{file=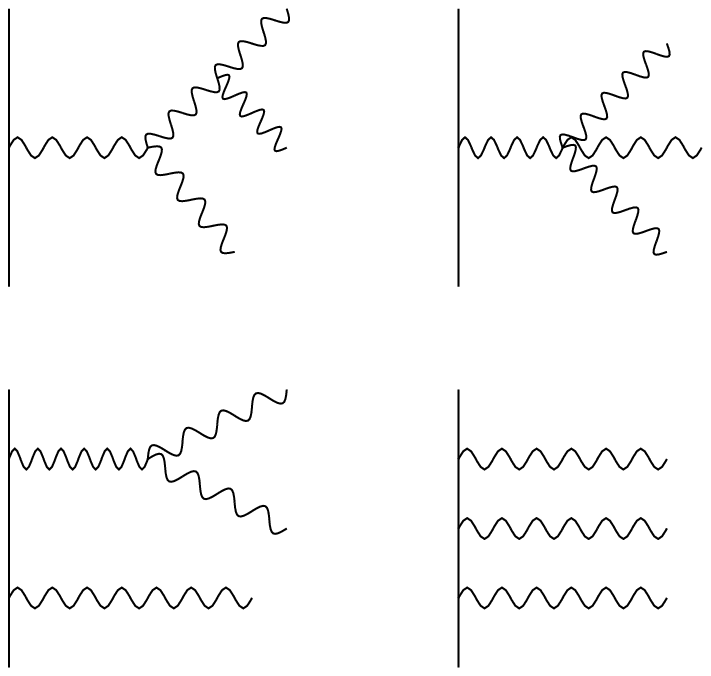,height=6.5cm,
bbllx=90pt,bblly=500pt,bburx=298pt,bbury=700pt}}
\caption{Diagrams contributing to ${\cal T}^{abc}_{\mu \nu \rho}$.
We do not show the diagrams which are related to those shown by exchange 
of outgoing bosons.}
\label{T3}
\end{figure}

Proceeding to the two-loop analysis, one must consider
two separate quantities: ${\cal A}_{2g}(s,\alpha^3)$, 
which is the contribution to the imaginary part of the two-loop
amplitude which arises from the convolution of the 
tree-level amplitude for $q(p_1)\bar{q}(p_2) \rightarrow g(k_1)g(k_2)$
with the hermitian conjugate of its one-loop partner, see 
Fig.\ref{T1}, and 
${\cal A}_{3g}(s,\alpha^3)$, which is the contribution to the imaginary part 
which arises on convoluting the tree-level amplitude for the process 
$q(p_1)\bar{q}(p_2) \rightarrow g(k_1)g(k_2)g(k_3)$ with its hermitian 
conjugate, see Fig.\ref{T3}. 
The two contributions must then be fed into a 
twice-subtracted dispersion relation and  
integrated from $4M^2$ to $\infty$ and from 
$9M^2$ to $\infty$ respectively, i.e.
\be
\Re {\rm e}\ \widehat{\Pi}(s,\alpha^2)) =
\frac{(s-M^2)^2}{\pi}
\Bigg[\int\limits^\infty_{4M^2}
\, \frac{ ds'\, \Im {\rm m} \ 
\widehat{\Pi}_{2g}(s',\alpha^2)}{(s'-M^2)^2 (s'-s)}\,
+\int\limits^\infty_{9M^2}
\, \frac{ ds'\, \Im {\rm m} \
\widehat{\Pi}_{3g}(s',\alpha^2)}{(s'-M^2)^2 (s'-s)}\Bigg]
\label{2LDR}
\ee
where
\begin{equation}
{\cal A}_{ng}(s,\alpha^3) = 2 D_0(s) \ V_{\mu c} V^{\mu}_c \ \Im {\rm m} \  
\widehat{\Pi}_{ng}(s,\alpha^2).
\end{equation}
If our calculations reveal that the RHS of (\ref{2LDR}) is infinite then
we will have shown that the 
$S$-matrix element for  $q \bar{q} \to q \bar{q}$
computed in the framework of the 
MYM is not renormalisable at two loops. 

We shall now show that ${\cal A}_{2g}$, when fed into the integral on the 
RHS of (\ref{2LDR}), gives a finite contribution, wheras the ${\cal A}_{3g}$ 
integral needs an additional subtraction in order to be rendered finite,
i.e. it gives rise to a non-renormalisable contribution.

To    see that the    contribution  from ${\cal  A}_{2g}$ contains  no
dangerous terms, it suffices to prove  that (a) the one-loop amplitude
${\cal    T}_{1\mu\nu}^{ab}$     
for $q \bar{q} \to gg$  
is     renormalisable,   and (b) that it satisfies exactly the
same type   of  BRST identity   as its  tree-level  counterpart ${\cal
  T}_{0\mu\nu}^{ab}$, i.e. that  (\ref{BRSg})   holds if  we   replace
${\cal  T}_{0\mu\nu}^{ab} \rightarrow    {\cal  T}_{1\mu\nu}^{ab}$ and
${\cal  S}_{0}^{ab}\rightarrow {\cal  S}_{1}^{ab}$.  Both (a)  and (b)
can be easily proved based on the analysis of \cite{BRS1loop}.  
It turns out that the closed
expressions for ${\cal  T}_{1\mu\nu}^{ab}$ and ${\cal S}_{1}^{ab}$ are
given by the Feynman diagrams of regular QCD in the Feynman gauge, but
with the tree-level propagators inside  all graphs replaced by massive
ones, again in the Feynman gauge, with the exception that for the ghost
contributions  we   have  a    different statistical    factor.  This
discrepancy does not affect the high energy  behaviour of such graphs,
i.e.  the ghost loops are well-behaved for large $q^2$. In addition,
as follows from \cite{BRS1loop}, the tree-level BRST identities do
indeed hold   at one loop. Thus    
\bea {\cal A}_{2g}(s,\alpha^3)  &=&
\frac{1}{2}\, \int 2 \  \Re   {\rm   e}   [ {\cal T}_{1\mu\nu}\,
Q^{\mu\rho}(k_1)\, Q^{\nu\sigma}(k_2)\, {\cal T}_{0\rho\sigma}^{\dagger}] 
d(P.S. {}^2)
\nonumber\\
&=&  \frac{1}{2}\,\int 2 \   \Re {\rm e}  [\Big( {\cal T}_{1}^{\mu\nu}
{\cal    T}_{0\mu\nu}^{\dagger}\    -\       {\cal S}_1{\cal S}_{0}^{\dagger}
\Big)]  d(P.S. {}^2) \eea 
and the  real  part can be
obtained using the twice-subtracted dispersion relation.

Now we turn to the amplitude
${\cal T}_{\mu\nu\rho}^{abc}(k_1,k_2,k_3)$ for the process
$q(p_1)\bar{q}(p_2) 
\rightarrow g(k_1)g(k_2)g(k_3)$, 
where $k_{i}$ is the four-momentum of the
$i$-th gluon, and $p_1+p_2=q=k_1+k_2+k_3$.
Such an amplitude is
given by the sum of the diagrams shown in Fig.\ref{T3} 

Let us compute the quantity 
\be
{\cal A}_{3g}(s,\alpha^3) = \frac{1}{3!}\int
[{\cal T}_{\mu\nu\rho}\,Q^{\mu\sigma}(k_1)\, 
\,Q^{\nu\lambda}(k_2)Q^{\rho\tau}(k_3)\,
{\cal T}_{\sigma\lambda\tau}^{\dagger}] d(P.S. {}^3)
\label{A3a}
\ee
where 
$\int d(P.S. {}^3)$ denotes the
integration over the three-body phase space, with the
combinatorial factor $1/3!$ accounting
for the three indistinguishable
gluons in the final state. As before, the polarisation tensors 
satisfy the transversality  condition: $k_{i} \cdot Q(k_i)=0$. 
At first sight, the integrand in (\ref{A3a}) seems to 
contain terms proportional 
to $(M^{-2})^{0}$, $(M^{-2})^{1}$, $(M^{-2})^{2}$ and $(M^{-2})^{3}$. 
The term proportional to $(M^{-2})^{0}$ is renormalisable, 
whereas all higher powers  give rise to non-renormalisable 
contributions: The higher the power, the worse the divergence.
However, as we shall shortly see, by virtue of the BRST
identities that ${\cal T}_{\mu\nu\rho}$ satisfies and the transversality 
properties of the polarisation tensors, only terms proportional to 
$(M^{-2})^{0}$ and $(M^{-2})^{1}$ survive. Thus, the worst divergence is 
logarithmic.

To establish this fact, let us first study the 
action of the longitudinal momenta $k_{1}^{\mu}$, $k_{2}^{\nu}$,
and $k_{3}^{\rho}$
on ${\cal T}_{\mu\nu\rho}$.
It is straightforward to verify that
${\cal T}_{\mu\nu\rho}^{abc}(k_1,k_2,k_3)$
satisfies the following identities:
\begin{eqnarray}
k^{\mu}_1{\cal T}_{\mu\nu\rho}^{abc} &=&
({\cal S} _{12})_{\rho}^{abc}k_{2\nu}
+ ({\cal S}_{13})_{\nu}^{abc}k_{3\rho} \, ,\nonumber\\
k^{\nu}_2{\cal T}_{\mu\nu\rho}^{abc} &=&
({\cal S}_{21})_{\rho}^{abc}k_{1\mu}
+({\cal S}_{23})_{\mu}^{abc}k_{3\rho} \, ,\nonumber\\
k^{\rho}_3{\cal T}_{\mu\nu\rho}^{abc} &=& 
({\cal S}_{31})_{\nu}^{abc}k_{1\mu}
+({\cal S}_{32})_{\mu}^{abc}k_{2\nu} \ .
\label{BRS3g}
\end{eqnarray}
Bose symmetry imposes the following relations among the
$S_{ij}$ amplitudes:
\bea
{\cal S}_{ij}^{a_{i}a_{j}a_{\ell}}(k_{i},k_{j},k_{\ell}) &=&
{\cal S}_{ji}^{a_{j}a_{i}a_{\ell}}(k_{j},k_{i},k_{\ell})
\nonumber\\
{\cal S}_{ij}^{a_{i}a_{j}a_{\ell}}(k_{i},k_{j},k_{\ell}) &=&
{\cal S}_{i\ell}^{a_{i}a_{\ell}a_{j}}(k_{i},k_{\ell},k_{j})
\label{Bose}  
\eea
and 
\be
k_{i}\cdot {{\cal S}}_{jl}= 
k_{j}\cdot {{\cal S}}_{il} 
~, ~~~l \neq i\neq j.
\label{Bose2}
\ee
The closed form of ${\cal S}_{23}^{abc}(k_{1},k_{2},k_{3})$ reads
\be
({\cal S}_{23})^{abc}_{\mu}=
({\cal S}^{s}_{23})^{abc}_{\mu}+({\cal S}^{t}_{23})^{abc}_{\mu}
\ee
with
\begin{eqnarray}
({\cal S}^{s}_{23})^{abc}_{\mu}&=&
g^2 V_{e}^{\alpha}\Bigg [f^{elc}f^{lab}
\frac{k_{3}^{\sigma}
\Gamma_{\alpha\sigma\mu}(q,k_{1}-q,-k_{1})}
{2k_2 \cdot k_3+M^2}
+f^{eal}f^{lbc}\frac{k_{3 \alpha}(k_1+k_2)_{\mu}}
{2k_1 \cdot k_2+M^2}
+f^{elb}f^{lac}\frac{k_{2 \alpha} k_{3 \mu} }{2 k_1 \cdot k_3 + M^2}
\Bigg]\nonumber \\
({\cal S}^{t}_{23})^{abc}_{\mu} &=& 
-ig^2 \bar{v}(p_2)
\left(
\, T^a \gamma_{\mu}\, 
\frac{1}{\not\! k_1 - \not\! p_2}
\, T^e \gamma_{\sigma}
+ T^e \gamma_{\sigma}\,
\frac{1}{\not\! p_1-\not\! k_1}\, 
\gamma_{\mu} T^a \, \right)  u(p_1)
\frac{f^{eac}k_{1}^{\sigma}}{2k_2 \cdot k_3+M^2}.
\nonumber\\
&&{} \label{S23}
\end{eqnarray}
In deriving the above expressions, in addition to the elementary Ward 
identity, (\ref{fundWI}), we have employed the tree-level Ward identity:
\bea
q^{\mu}_{1}\Gamma_{\mu\nu\alpha\beta}^{abcd}(q_1,q_2,q_3,q_4) =&&
f^{abe}\Gamma_{\alpha\beta\nu}^{cde}(q_3,q_4,q_1+q_2)
+f^{ace}\Gamma_{\beta\nu\alpha}^{dbe}(q_4,q_2,q_1+q_3)\nonumber\\
&&+ f^{ade}\Gamma_{\nu\alpha\beta}^{bce}(q_2,q_3,q_1+q_4), 
\eea
which relates the bare three- and four-gluon vertices.
All remaining $S_{ij}$ amplitudes can be obtained from $S_{12}$
using the relations of (\ref{Bose}).

We next let the longitudinal momenta in the
polarisation tensors act on  ${\cal T}_{\mu\nu\rho}$ and 
${\cal T}_{\sigma\lambda\tau}$, and use  (\ref{BRS3g}). We can see how 
the $(M^{-2})^{3}$ terms disappear. The action of the term 
$M^{-2}k_{3}^{\rho} k_{3}^{\tau}$ gives terms proportional to 
$k_{1}^{\mu}$ and $k_{2}^{\nu}$ (or equivalently $k_{1}^{\sigma}$ 
and $k_{2}^{\lambda}$), which vanish when they hit  
$Q^{\mu\sigma}(k_1)$ or $Q^{\nu\lambda}(k_2)$. So, (\ref{A3a}) reduces to
\be
{\cal A}_{3g} = \frac{1}{3!}\int
\Bigg[{\cal T}_{\mu\nu\rho}Q^{\mu\sigma}(k_1)\,
Q^{\nu\lambda}(k_2)\,{\cal T}_{\sigma\lambda\rho}^{\dagger}\Bigg]
d(P.S. {}^3).
\label{A3b}
\ee
At this point, the highest possible power of $M^{-2}$
is $(M^{-2})^{2}$ . We now let
$M^{-2}k_{2}^{\nu} k_{2}^{\lambda}$ act on ${\cal T}_{\mu\nu\rho}$
and ${\cal T}_{\sigma\lambda\rho}^{\dagger}$:
\bea
M^{-2}k_{2}^{\nu} k_{2}^{\lambda}{\cal T}_{\mu\nu\rho}\,
Q^{\mu\sigma}(k_1)\,{\cal T}_{\sigma\lambda\rho}^{\dagger}
&=&
({\cal S}_{21}^{\rho}k_1^{\mu}
+{\cal S}_{23}^{\mu}k_3^{\rho})\,Q_{\mu\sigma}(k_1)\,
({\cal S}_{21 \rho} k_1^{\sigma} + {\cal S}_{23}^{\sigma}k_{3\rho})^{\dagger}
\nonumber\\
&=&{\cal S}_{23}^{\mu}{\cal S}_{23}^{\sigma \dagger}\,Q_{\mu\sigma}(k_1).
\label{A3c}
\eea
Thus, one more power of $M^{-2}$ has been eliminated. We are left with
\bea
{\cal A}_{3g} &=& \frac{1}{3!}\int
[({\cal T}_{\mu\nu\rho}{\cal T}_{\sigma\nu\rho}^{\dagger}
-{\cal S}_{23}^{\mu}{\cal S}_{23}^{\dagger \sigma})\,Q^{\mu\sigma}(k_1)]
d(P.S. {}^3),
\nonumber\\
&=&{\cal A}_{R}+{\cal A}_{NR},
\label{A3d}
\eea
where
\be
\label{Ren}
{\cal A}_{3g}^{R} = \frac{1}{3!}\int
[{\cal T}_{\mu\nu\rho}{\cal T}_{\mu\nu\rho}^{\dagger}
- 3({\cal S}_{12}{\cal S}_{12}^{\dagger})]d(P.S. {}^3),
\ee
\be
\label{NoRen}
{\cal A}_{3g}^{NR} =-\frac{1}{3!}\frac{1}{M^{2}} \int
(k_1 \cdot {\cal S}_{23})(k_1 \cdot {\cal S}_{23})^{\dagger} d(P.S. {}^3).
\ee
In deriving the above expressions we have used the fact that
the phase-space integration is invariant under 
$k_{i} \leftrightarrow k_{j}$. 
 The term in (\ref{NoRen}) will generate
non-renormalisable terms; it is clearly non-vanishing,
since it is a three-body phase-space integral over a positive 
definite quantity. In particular,
\be
{\cal A}_{3g}^{NR}= 
-\frac{1}{3!}\frac{c_A^2}{8M^2} \int
V^{e}_{\alpha}(k_1-k_2)^{\alpha}(k_1-k_2)^{\beta}
V^{e}_{\beta}d(P.S. {}^3) \label{nonrenorm}
\ee
where the
identities $f^{abe}f^{cde}+f^{ace}f^{dbe}+f^{ade}f^{bce}=0$
and $f^{alm}f^{bmn}f^{cnl}=\frac{1}{2}c_A f^{abc}$ have been used. We also
used $\frac{k_{i} \cdot k_{j}}{2k_{i}k_{j}+M^{2}}= \frac{1}{2}+ ...$,
where the omitted term is proportional to $M^{2}$, thus giving rise to a
renormalisable contribution, i.e. the omitted term belongs effectively 
to $ {\cal A}_{3g}^R$. Notice that the non-renormalisable terms are 
{\it purely propagator-like} (universal, process-independent).

Finally, it is instructive
to compare the result of (\ref{A3d}) with that of normal QCD.
In the QCD case there are, of course, no terms proportional to $M^{-2}$ 
or higher powers. On the other hand, the presence of the auxiliary
four-vector $\eta_{\mu}$ in the polarisation tensors could in principle induce
spurious divergences, should it survive in the final answer. It is
easy to see however how any reference to $\eta_{\mu}$ disappears before any of 
the phase-space integrations are carried out. 
Let us denote the corresponding QCD amplitude by
${\widehat {\cal T}}_{\mu\nu\rho}$.
We start again with 
\be
{\cal A}_{3g}^{QCD} = \frac{1}{3!}\int
[{\widehat {\cal T}}_{\mu\nu\rho}\,
\,P^{\mu\sigma}(k_1)\, \,P^{\nu\lambda}(k_2)\,P^{\rho\tau}(k_3)\,
{\widehat {\cal T}}_{\sigma\lambda\tau}^{\dagger}]
d(P.S. {}^3)
\label{A3QCDa}
\ee
where the gluons are now massless and $k_{i} \cdot P(k_i)=0$ ~$(i=1,2,3)$.  
Equation (\ref{BRS3g}) is valid
for QCD, as can be shown rigorously using BRST arguments. 
Since the elementary Ward identity (\ref{fundWI}) is the same for
both MYM and massless QCD, it follows that the closed expressions for  
the factors $S_{ij}$ in (\ref{BRS3g}) may be recovered from 
(\ref{S23}) simply by setting $M^2=0$. We
denote them by ${\widehat{S}}_{ij}$. Then, by letting the longitudinal 
momenta act on ${\widehat {\cal T}}_{\mu\nu\rho}$ and using 
(\ref{BRS3g}), one can easily verify that any reference to
the four-vector $\eta_{\mu}$ disappears and that the final answer is
\be
{\cal A}_{3g}^{QCD} = \frac{1}{3!}\int
[{\widehat {\cal T}}^{\mu\nu\rho}
{\widehat {\cal T}}_{\mu\nu\rho}^{\dagger} - 
6{\widehat{{\cal S}}}_{12}{\widehat{{\cal S}}}_{12}^{\dagger}] d(P.S. {}^3).
\ee
So, unlike the MYM model, in massless QCD all potentially dangerous terms 
vanish. 

\setcounter{equation}{0}
\section{Connection to field theories with a Higgs mechanism}
\protect\label{sec:Higgs} 

It  is well known  that the only way  to endow  gauge fields with mass
whilst maintaining unitarity and  renormalisability  is via the  Higgs
mechanism \cite{CLT,CLS,LQT}.  This  procedure is not suitable however
for an  effective model of strong  interactions because  it introduces
extra  scalar particles in  the physical spectrum.    In the MYM model
massive  gauge fields are  obtained without introducing extra physical
fields, at the  price of losing  renormalisability at higher orders of
perturbation theory.  It is instructive to see explicitly how the lack
of renormalisability in the MYM can be traced back to the absence of a
Higgs particle; in  particular understand why the process $q\bar{q}\to
q\bar{q}$ (and  $q\bar{q}\to gg$) is  renormalisable  at one-loop, but
ceases   to  be renormalisable beyond  one-loop.    In this section we
address these issues in  detail by performing a quantitative  analysis
of the differences and similarities between the  MYM and a Higgs model
(HM)  at the  level  of  the $S$-matrix.   In addition,   as has  been
discussed in detail in \cite{Cornwall},  it is possible to speak about
the MYM using  the language of a HM.  Specifically, one can think of
the MYM as a theory where all gluons have been  given masses by adding
to the Lagrangian   a   sufficient number  of  Higgs  multiplets  ($N$
fundamental representations of  $SU(N)$ ),  and then ``freezing''  all
the polar excitations of the  $N$ Higgs fields.  The remaining $N^2-1$
angular  excitations   corresponding  to  the  Goldstone   bosons  are
precisely      the    angular    fields $\theta^{a}$    displayed   in
(\ref{angular}).

To illustrate the above points we will 
use  a toy Higgs field theory which
displays all the essential features we want to study.
The gauge group
of this model is SU(2).  The Higgs mechanism  is triggered by a
complex doublet $\phi$  in   the fundamental representation  (isospin
$l=\frac{1}{2}$).  This particular  assignment endows all  three gauge
bosons with the same mass $M$, whilst simultaneously prohibiting terms
of the form ${\phi}\bar  {\psi}{\psi}$ for any fermion  representation
of isospin  $\ell$ \footnote{It is  elementary to verify that no gauge
  singlet  (total $\ell=0$)  can be  formed out  of the  above isospin
  assignments.}.    To   mimic   QCD,  we    choose  the   fundamental
representation for the massless  fermion fields $\psi$,  although this
choice is    not  essential  for  what   follows.   As  there   are no
interactions  between   fermions and    scalars, the   fermions remain
massless even when the   scalar fields acquire a non-vanishing  vacuum
expectation value.  The Lagrangian density for this model is 
\be {\cal
  L}=      -\frac{1}{4}{\cal  F}_{\mu\nu}{\cal    F}^{\mu\nu}   + \bar
{\psi}{\not\!                D}                 {\psi}               +
({D}_{\mu}{\phi})({D}^{\mu}{\phi})^{\dagger}      -
V({\phi})  \ee  with \bea    ({\not\!  D}{\psi})_{\alpha} &=&{\not\!   
  \partial}{\psi}_{\alpha}
+ igT^{a}_{\alpha\beta}{\psi}_{\beta}{\not\! A}^{a}~,\nonumber\\
(D_{\mu}{\phi})^{i} &=&\partial_{\mu}{\phi}^{i} +
ig{T}^{ij}_{a}{\phi}_{j}A^{a}_{\mu}~,\nonumber\\
V(\phi)            &=&          {\mu}^2{\phi}{\phi}^{\dagger}
+\lambda({\phi}{\phi}^{\dagger})^{2},        \eea        where
${T}_{a}=\frac{1}{2}{\sigma}_{a}$,   and ${\sigma}_{a}$  are the Pauli
matrices.  If   ${\mu}^2<0$, the Higgs mechanism   gives rise to three
degenerate massive   gauge bosons  of    mass  $M= gv/2$,  where   $v=
\sqrt{-\frac{\mu^2}{\lambda}}$ is the minimum of $V(\phi)$.  The above
model is a vector-like variant of the usual electroweak
 sector of the
Standard Model, $SU(2)\times U(1)$, with the Weinberg angle $\theta_W$
set to zero.

The corresponding bare gauge boson propagator in the $R_{\xi}$ gauge
 has the form  
\be
\Delta_{\mu\nu}=\Bigg ( g_{\mu\nu}-\frac{q_{\mu}q_{\nu}(1-\xi)}
{q^2-{\xi}M^2}\Bigg )\frac{-i}{q^2-M^2}
\ee
and the would-be Goldstone boson ($G$) and ghost ($c$) propagators are
\be
\Delta_{G,c}=\frac{i}{q^2-{\xi}M^2}.
\ee
In the unitary gauge, which formally corresponds
to the limit $\xi\rightarrow\infty$, the gauge boson propagator
takes the form (\ref{uniprop}) and there are no Goldstone 
boson/ghost propagators.
\footnote{The
renormalisability of the  HM in the unitary  gauge is not manifest. 
For example,  it is known that,   even though  the
$n$-point functions  are  non-renormalisable,  by  virtue  of  subtle
cancellations,  the  $S$-matrix  element built   out of  these
non-renormalisable  $n$-point functions can be  made  finite with the
usual mass  and charge renormalisation \cite{HMren}.  A  
more immediate way  to see this is to resort from  the beginning of  
the calculation to the PT rearrangement of the
amplitude \cite{PS}.}
Finally, there is a Higgs particle of mass $M_{H}= v\sqrt{2\lambda}$ with
bare propagator 
\be
\Delta_{H}=\frac{i}{q^2-M_{H}^2} ~.
\ee
The one-loop $\beta$-function for the gauge coupling has the form \cite{GW}
\be
\beta = -   \frac{1}{16{\pi}^2} (b_{g}-b_{f}-b_{s}) g^3
\ee
with
\be
   b_{g}=\frac{11}{3}c_{A}~, ~~~
   b_{f}=\frac{4}{3}n_{f}T_{f}~,~~~ 
   b_{s}=\frac{1}{6}n_{s}T_{s}~,
\label{betacoeff} 
\ee
where $T_{f}$ is the Dynkin index of the fermion representation,
$T_{s}$ is the Dynkin index of the scalar representation,
$n_{f}$ is the number of fermion families in a given representation
and $n_{s}$ the number of real scalar families. For the 
particular scalar representation we have chosen, $T_{s}=\frac{1}{2}$ and
$n_{s}=2$. In the absence of quarks we have that 
$\beta=-\frac{g^3}{16{\pi}^2}(43/6)$.

\begin{figure}[t] 
\centerline{\epsfig{file=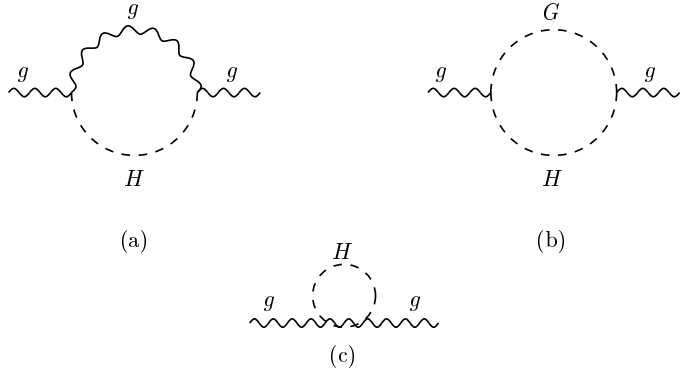,height=5cm,
bbllx=137pt,bblly=472pt,bburx=463pt,bbury=652pt}}
\caption{Higgs boson contributions to $\widehat{\Pi}_{\mu \nu}^{gH}$.
In the unitary gauge recall that (b) is absent.}
\label{Higgs}
\end{figure}

Let us now proceed with a study of the one-loop amplitude.
As we already showed, at the level of $S$-matrix elements the
MYM and the naive model are equivalent. In addition, if we adopt the
unitary gauge for the HM then it is obvious that, 
to any finite order in perturbation theory,
the only difference between an $S$-matrix element computed 
in the MYM and the corresponding $S$-matrix element computed in the HM
is due to contributions to the latter which come from Feynman 
diagrams containing Higgs boson propagators \cite{HeavyHiggs}. 
For example, in the case of one-loop quark scattering,
in addition to the graphs in Fig.\ref{loop}, which are common
to both MYM and HM, the diagram of Fig.\ref{Higgs}(a) contributes to the 
$S$-matrix element of the HM. 
In other words, the $S$-matrix elements of the MYM model may be obtained
from the corresponding $S$-matrix elements of HM by
omitting all diagrams containing a Higgs 
particle. 
Given this observation, it is easy to see why the 
the process $qq\to qq$ in the 
MYM is renormalisable
at one loop: The only difference between the renormalisable HM 
and the MYM is the contribution corresponding to
the graphs of Fig.\ref{Higgs}, which themselves form a gauge-invariant 
and renormalisable subset. 
Denoting their contribution by $\widehat{\Pi}_{\mu\nu}^{gH}$ 
we have that, up to the immaterial tadpole graph,
\be
\widehat{\Pi}_{\mu\nu}^{gH}= g^2 M^2
\int\frac{d^nk}{i(2\pi)^n}
\left(- g_{\mu\nu}+ \frac{k_{\mu}k_{\nu}}{M^2} \right)
\frac{1}{(k^2-M^2)[(k+q)^2 - M_H^2]}.
\ee
The factor $M^2$ in front of
the integral originates from the gluon-gluon-Higgs coupling and 
guarantees that 
$\widehat{\Pi}_{\mu\nu}^{gH}$ can be made ultraviolet-finite
by means of the usual mass and wave function renormalisation.
After (on-shell) renormalisation, the above expression
becomes, in the limit $q^2\gg M^2$, (and dropping the terms 
proportional to $q^{\mu}q^{\nu}$)
\be
\widehat{\Pi}_{\mu\nu}^{gH}(q^2)=
-\frac{\alpha}{4\pi} 
\left( \frac{1}{12} \right) q^2\ln(q^2/M^2)g^{\mu\nu} + ...
\label{gHLimit}
\ee where the ellipsis denotes  numerical constants and terms of order
$O(M^2/q^2)$.   If  we   now set   $c_A=2$    in  the expression    of
(\ref{PR2Limit}), and add it to the  expression in (\ref{gHLimit}), we
see that the coefficient in front of the resulting logarithmic term is
equal to  $43/6$,  which is  precisely  the   coefficient of  the   HM
$\beta$-function without quarks.  In  addition, as  expected  from the
discussion on the  connection between the MYM and  the HM given at the
beginning  of this  section,   the expression  in (\ref{gHLimit})   is
exactly the difference between (4.27) and (4.28) for $c_A=2$.

\begin{figure}[t] 
\centerline{\epsfig{file=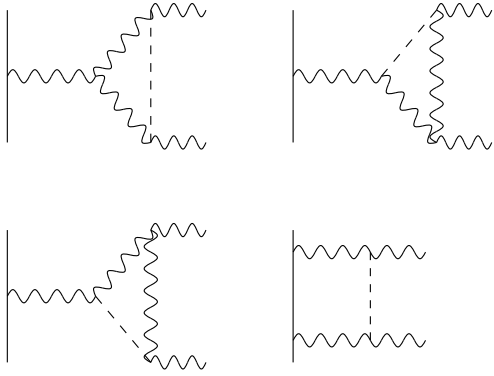,height=5cm,
bbllx=88pt,bblly=523pt,bburx=318pt,bbury=694pt}}
\caption{The Higgs graphs contributing to ${\cal T}_1^H$. The graphs of
Fig.\ref{T1} must also be included when computing ${\cal T}_1^H$.}
\label{Higgs2}
\end{figure}

We now turn to the two-loop analysis. Firstly,
it is relatively straightforward to establish that the contribution from the
two-gluon cut of those one-loop diagrams which contain a Higgs boson,
Fig.\ref{Higgs2}, gives rise to a renormalisable contribution, i.e. the 
corresponding dispersive (real) part can be
made finite by means of a twice-subtracted dispersion relation.
This is of course expected, since the one-loop diagrams
contributing to ${\cal A}_{2g}$ which we studied 
in the previous section were themselves renormalisable,
i.e. no cancellation from diagrams containing
a Higgs boson is required.

To see this explicitly, consider the amplitude 
${\cal A}_{2g}^{H}$ given by 
\be
{\cal A}_{2g}^{H}(s) = \frac{1}{2}\, \int 2 \ \Re {\rm e}
[ {\cal T}_{1\mu\nu}^{H}\, 
Q^{\mu\rho}(k_1)\, Q^{\nu\sigma}(k_2)\, 
{\cal T}_{0\rho\sigma}^{\dagger}] d(P.S. {}^2)
\ee
where ${\cal T}_{1\mu\nu}^{H}$ is shown in Fig.\ref{Higgs2}. 
It is straightforward to verify that 
${\cal T}_{1\mu\nu}^{H}$ is a gauge-independent 
quantity, and that it satisfies
\be
k^{\mu}_1{\cal T}_{1\mu\nu}^{H} = k_{2\nu}{\cal S}_{1}^{H}
\label{BRSH}
\ee
with
\be
{\cal S}_{1}^{H} = gV^{\lambda}D^0(q)
[\widehat{\Pi}^{gH}_{\lambda\alpha}D^0(q)](k_1-k_2)^{\alpha}.
\label{SH}
\ee
Notice that the expression in square brackets behaves like
$\log(s/M^2)$ for $s\gg M^2$. Using (\ref{BRSg}) and   (\ref{BRSH})
we can see that
\be
{\cal A}_{2g}^{H}(s)=
\frac{1}{2}\,\int 2 \Re {\rm e}
[\Big( {\cal T}_{1\mu\nu}^{H} {\cal T}_{0}^{\mu\nu\dagger}\ -\ 
{\cal S}_1^{H}{\cal S}_{0}^{\dagger}\Big)] d(P.S. {}^2).
\label{AH}
\ee
So, this contribution gives rise to renormalisable two-loop
amplitudes.

\begin{figure}[t] 
\centerline{\epsfig{file=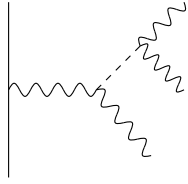,height=2.5cm,
bbllx=70pt,bblly=594pt,bburx=155pt,bbury=680pt}}
\caption{One of the graphs contributing to ${\cal T}_{\mu \nu \rho}^H$. The
others are obtained by permuting the outgoing bosons.}
\label{Higgs3}
\end{figure}

To see how the presence of the Higgs boson enforces 
 renormalisability,
we focus on the two amplitudes, ${\cal A}_{3g}^{H,a}$
and  ${\cal A}_{3g}^{H,b}$:
\be
{\cal A}_{3g}^{H,a}(s) = \frac{1}{3!}\int
[{\cal T}_{\mu\nu\rho}^{H}\,Q^{\mu\sigma}(k_1)\, 
\,Q^{\nu\lambda}(k_2)Q^{\rho\tau}(k_3)\,
{\cal T}_{\sigma\lambda\tau}^{H\dagger}] d(P.S. {}^3)
\label{A3gHa}
\ee
and
\be
{\cal A}_{3g}^{H,b}=
\frac{1}{3!}\int
2\Re {\rm e}
[{\cal T}_{\mu\nu\rho}^{H}\,Q^{\mu\sigma}(k_1)\, 
\,Q^{\nu\lambda}(k_2)Q^{\rho\tau}(k_3)\,
{\cal T}_{\sigma\lambda\tau}^{\dagger}] d(P.S. {}^3)
\label{A3gHb}
\ee
where the amplitude ${\cal T}_{\mu\nu\rho}^{H}$ is shown in Fig.\ref{Higgs3}.
As can be seen, ${\cal A}_{3g}^{H,a}$ arises by multiplying only those
three-gluon amplitudes which contain Higgs particles whilst  
${\cal A}_{3g}^{H,b}$ comes from interfering the Higgs diagrams with the
non-Higgs diagrams of Fig.\ref{T3}. Since the coupling of the Higgs boson to
two gauge bosons is proportional to $M$, it follows that 
${\cal T}_{\mu\nu\rho}^{H}$ has already a factor $M^2$
built into it. Consequently, there is an implicit factor $M^4$
inside ${\cal A}_{3g}^{H,a}(s)$, and therefore 
the only non-renormalisable
contribution in ${\cal A}_{3g}^{H,a}(s)$ will come from the
term in the polarisation tensors which is proportional
to $(M^{-2})^3$. We therefore find that
\be
[{\cal A}_{3g}^{H,a}]_{NR}(s)=
-(\frac{1}{2})\frac{1}{3!}\frac{1}{M^2} \int
V^{e}_{\alpha}(k_1-k_2)^{\alpha}(k_1-k_2)^{\beta}
V^{e}_{\beta}d(P.S. {}^3).
\ee
The interference term, ${\cal A}_{3g}^{H,b}$, has an implicit $M^2$ inside so
now contributions from the $(M^{-2})^3$ and $(M^{-2})^2$ terms in the 
polarisation tensors are needed. One finds, using  (\ref{BRS3g}),
the following non-renormalisable contribution:
\be
[{\cal A}_{3g}^{H,b}]_{NR}(s)=
\frac{1}{3!}\frac{1}{M^2} \int
V^{e}_{\alpha}(k_1-k_2)^{\alpha}(k_1-k_2)^{\beta}
V^{e}_{\beta}d(P.S. {}^3).
\ee
In arriving at the above results, 
identities of the type
$k_1^{\alpha}k_1^{\beta}= \frac{1}{3}(k_1-k_2)^{\alpha}(k_1-k_2)^{\beta}$,
or $k_2^{\alpha}k_3^{\beta}=-\frac{1}{2}k_3^{\alpha}k_3^{\beta}$
valid under the integral sign, may be found useful.
Comparing to (\ref{nonrenorm}) of the previous section (setting $c_A=2$), 
we see that the Higgs contribution exactly cancels the non-renormalisable
part of the MYM two-loop contribution. 
Evidently, even though the Higgs boson does not couple directly to
the quarks,  (since in this toy
model the gauge
symmetry prohibits Yukawa couplings)
its importance in restoring the renormalisability
of the process $q\bar{q}\to q\bar{q}$
manifests itself through
the tree-level sub-amplitudes 
$gg\to gg$ containing the Higgs boson
(Fig.7), which 
reside in the two-loop
diagrams.

\section{Quark-Quark Elastic Scattering}
\protect\label{sec:pomeron}
In this section we take an introductory look at the elastic scattering of a 
pair of quarks via two-gluon exchange within the MYM model. 
Of course, quark-quark elastic scattering cannot be measured directly, but it 
is possible that many of the elements which are central to the more realistic 
processes (e.g. hadron-hadron elastic scattering) are contained in this 
simpler treatment. This is in the spirit of the Donnachie-Landshoff-Nachtmann
approach \cite{LN,DL}, where the success of the additive quark rule provides 
evidence that one need not know about the detailed structure of the 
colliding hadrons before one can proceed to make elastic scattering 
calculations, although it is not yet established that this is 
correct \cite{BP}.

As a first step, one can calculate the amplitude for the elastic scattering of 
differently flavoured quarks, i.e. $q_i q_j \to q_i q_j$,
at the lowest order, keeping only those terms which dominate in the 
Regge limit. This is a straightforward calculation of the box diagram shown in
Fig.\ref{loop}(c) rotated through 90 degrees 
(the crossed box diagram contributes only to the real part 
of the amplitude in the Regge limit and constitutes a sub-leading correction).
The leading contribution is imaginary and so can be obtained directly using 
the cutting rules, i.e. the amplitude for single gluon exchange can be 
written:
\begin{equation}
A(s,t)_{1-{\rm gluon}} = 
i T^a \otimes T^a 2 p_1^{\mu} \frac{g_{\mu \nu}}{k^2-M^2} 2 p_2^{\nu}
\delta_{\lambda_1 \lambda_1'} \delta_{\lambda_2 \lambda_2'},
\end{equation}
where $k$ is the momentum of the exchanged gluon and $p_1$ and $p_2$ are
the momenta of the incoming quarks, i.e. $s = (p_1+p_2)^2$.
The high-energy limit allows the exchanged gluon to be assumed soft, and
so the eikonal approximation has been used to simplify the $qqg$ vertex. 
The delta functions ensure helicity conservation at each vertex. 
Multiplying by the conjugate amplitude, projecting out the colour singlet part
and performing the two-body phase space integral 
(putting the intermediate quarks on-shell) allows us to write
\begin{equation}
A(s,t) = i s \alpha_s^2 \frac{N^2-1}{N^2} 
\int d^2 {\bf k} \frac{1}{({\bf k}^2+M^2)(({\bf k-q})^2+M^2)},
\end{equation}
where ${\bf q}^2 = -t > 0$,  $N$ is the number of colours
and the exchanged gluons are taken to be purely transverse. 
The transverse momentum integral can be performed and yields
\begin{equation}
A(s,t) = i \frac{s}{-t} 2 \pi \alpha_s^2 \frac{N^2-1}{N^2} \frac{1}{\Delta(t)}
\ln \frac{\Delta(t)+1}{\Delta(t)-1},
\end{equation}
where $\Delta(t)$ is defined in (\ref{delta}).
Thus, the total cross-section for $q_i q_j \to X$ is
\begin{equation}
\sigma_T = \frac{1}{M^2} \pi \alpha_s^2 \frac{N^2-1}{N^2}.
\end{equation}

It is instructive to investigate the conditions under which the 
two-gluon exchange amplitude calculated above violates unitarity. 
We shall see that unitarity is violated only for very central collisions 
and that these constitute an insignificant fraction of the total and 
elastic scattering cross-sections. Only for very high-$t$ processes do 
we have collisions
which are sufficiently central to cause a worry. This gives us confidence
to proceed to the next order of calculation, assured that we have yet to
receive indications that unitarisation corrections are important. 

To investigate unitarity we perform a Fourier transform of the elastic
scattering amplitude, i.e.
\begin{equation}
\tilde{A}(s,b) = 
\int \frac{d^2 {\bf q}}{(2 \pi)^2} e^{-i {\bf q} \cdot {\bf b}} 
\frac{A(s,t)}{2 s}, \label{fourier}
\end{equation}
and ${\bf b}$ is the impact parameter of the collision. Written in this
way, unitarity demands that $$ | \tilde{A}(s,b) |^2 < 1 $$ 
for all ${\bf b}$. However, we can be confident that
unitarisation corrections are small if the inequality is satisfied for those
values of impact parameter which dominate the process under study.
Numerical evaluation of  (\ref{fourier}) demonstrates that the amplitude
only ever violates unitarity for $Mb < 10^{-2}$, $10^{-3}$  for 
$2/\alpha_s^2 = 20$, $50$ respectively, i.e. only for very central collisions
(on the scale of the gluon mass). In this language, the total cross-section 
is given by
\begin{equation}
\sigma_T = 2 \int d^2 {\bf b}\  \tilde{A}(s,b)
\end{equation}
whilst the elastic scattering cross-section is given by
\begin{equation}
\sigma_{{\rm el}} = \int d^2 {\bf b} \ |\tilde{A}(s,b)|^2.
\end{equation} 
Since $\tilde{A}(s,b)$ decreases monotonically as $Mb$ increases, 
it follows that
the elastic scattering cross-section receives a larger contribution from
more central collisions than the total cross-section. To a first 
approximation, the typical impact parameter is set by the gluon mass, i.e.
$\langle b^2 \rangle = C/M^2$ where $C \sim 1$ and is larger for the total
cross-section than for the elastic cross-section. In either case, we
are always well away from the dangerous region where unitarity is violated. The
situation is different for high-$t$ processes, since now 
$\langle b^2 \rangle \sim 1/|t|$ and so, for $\sqrt{-t} > (10^2 - 10^3) M$
we would need to worry that unitarisation corrections are important. 

\section{Conclusion and Perspectives}
\protect\label{sec:conclusions}
In this paper we have reviewed 
and investigated the formalism of the MYM model, arguing
that it  may   be  relevant as a tool to investigate diffractive scattering 
(and possibly other areas of strong interactions phenomenology), where 
traditional QCD methods are inadequate. 
A detailed study of the $q\bar{q} \to q\bar{q}$ process in the context
of  this model   up  to the two-loop  order   was presented,  and  the
renormalisation properties of the corresponding $S$-matrix  
were discussed.  

 Let us summarise briefly the prospects for a study of diffractive scattering. 
A successful model  of
diffraction should   be able  to  explain: The growth  of  total
hadronic cross-sections with increasing $s$.  In particular, the model 
 should show why the rise  in soft hadronic processes (e.g. the
total $pp$ cross-section) proceeds at a much  slower rate than in hard
processes (e.g. the $\gamma^*  p$ cross-section);  the shrinkage of
the forward  diffraction peak with  increasing $s$. 
In other words, the model should  be able  to explain  the  
qualitative  success of the
Donnachie-Landshoff-Nachtmann model of the  pomeron as a  single Regge
pole in soft  diffraction (i.e. in  those processes where there  is no
large scale) and its  failure in small-$x$ deep inelastic  scattering,
in the diffractive production of  all vector mesons at  high $Q^2$ and  in
the diffractive production  of $J/\Psi$ mesons  at low $Q^2$.   

In the
future, we plan to  use the MYM model to  compute the complete  ${\cal
  O}(\alpha_s)$    corrections  to the   two-gluon-exchange amplitude
discussed in the previous section, in order  to verify whether some or
all of the aforementioned features emerge.
More specifically,  such a calculation  should help us investigate the
following points:
\begin{itemize}
\item In the limit of large enough $s$, the logarithms $\sim (\alpha_s
  \ln s)^m$, which appear at each  order in perturbation theory, become
  large  and it  becomes  necessary to sum them   to all orders.  This
  summation of leading logarithms  is performed using the formalism of
  BFKL. It  is an open  question  as to precisely  when this summation
  leads to the dominant contribution  to the amplitude. In fact, since
  the  summation   is of leading  logarithms   only,  we cannot define
  exactly  what it means to  say $s$ is large,  although we note that, 
  in this respect,  analysis of the next-to-leading   
  logarithmic corrections calculated  by
  Fadin,  Lipatov et al. \cite{NLO} should  improve  the situation. 
By
  computing  at fixed order  in   $\alpha_s$  we can  investigate  the
  relative importance of the $\sim  \alpha_s \ln  s$ term compared  to
  the terms  which do not  include the logarithm.  In this way, we can
  make some quantitative statements regarding the need (within the MYM
  model)  to sum the  remaining  leading  logarithms. For example,  it
  might   be  that, at  the  energies  of  contemporary colliders, the
  logarithm is not so large to justify dropping the  other terms, i.e. 
  a fixed order calculation might be the better way to proceed.

It is known that 
 introducing a gluon mass has a very small effect on the leading logarithmic
contribution \cite{diffraction1,BFKL}. This arises largely because
the BFKL summation is infrared finite, i.e. there are infrared 
cancellations between real and virtual graphs which reduce the sensitivity 
to this region. These cancellations persist even after adding a gluon mass
(via a Higgs, or via the MYM model) and 
 serve to reduce the sensitivity of the 
amplitudes to variations in the mass. 
Note that is not too important how
the mass is introduced. This can
be seen since no Higgs graphs contribute to the leading logarithm summation
and since the gauge-dependent part of the gluon propagator is also sub-leading
(in covariant gauges).

\item It is also known that the leading logarithm summation leads
to a rapid rise of total cross-sections. It can be argued that this
rapid rise, which is due to multiple soft gluon emission, reveals itself in 
hard scattering processes, but is masked in softer processes by unitarity
corrections. 
Any slowing down of the rise via unitarity corrections has
yet to be precisely quantified. Another possibility is that the 
strong rise seen in hard processes can be explained in fixed-order 
perturbation theory, i.e. arising from the $\ln s$ term, and that 
this same rise is masked in soft processes by a non-logarithmic contribution 
which is comparable in size to the logarithmic contribution (i.e. 
as the process becomes harder, the non-logarithmic contribution falls away 
to reveal the logarithm). This latter possibility can be investigated 
after computing the radiative corrections to the two-gluon exchange graphs. 
\end{itemize}
In summary, we think that the MYM may prove a useful tool in understanding 
the phenomenology of diffractive scattering by bridging the gap between 
different QCD-inspired models. Such a conjecture will be tested through 
next-to-leading order calculations of quark-quark elastic scattering, which 
we plan to discuss in a forthcoming paper.

\bigskip

\vspace{0.7cm}\noindent {\bf    Acknowledgments}      
We thank V. Del Duca and M. Testa for some useful discussions.
This work was supported in part by the EU Fourth Programme `Training and 
Mobility of Researchers', Network `Quantum Chromodynamics and the Deep 
Structure of Elementary Particles', contract FMRX-CT98-0194 (DG 12-MIHT). 
The work of JP is funded by a Marie Curie Fellowship (TMR-ERBFMBICT 972024). 
JP also acknowledges financial support from the Department of Physics and 
Astronomy of the University of  Manchester while parts of this work
were being completed. CP acknowledges the support of PPARC through an 
Advanced Fellowship.

\end{document}